%% file: main.tex
\documentclass[preprint, 11pt]{elsarticle}
\usepackage[margin=1in]{geometry}
\usepackage[utf8]{inputenc}
\usepackage{cmap}
\usepackage{textcomp}
\usepackage{microtype}
\usepackage{subfiles}
\usepackage[numbers]{natbib}
\usepackage{mathtools}
\usepackage{graphicx}
\usepackage{xurl}
\usepackage{varioref}
\usepackage{hyperref}
\usepackage{cleveref}
\usepackage{xcolor}
\usepackage{color,soul}
\usepackage{siunitx}
\usepackage{float}
\usepackage{tikz}
\usepackage{pgfgantt}
\usepackage{pgfplots}
\usepackage{multirow}
\usepackage{bm} 
\usepackage{multicol}
\usepackage{booktabs}
\usepackage{subcaption}
\usepackage{amsmath}
\usepackage{amsopn}
\usepackage{amssymb}
\usepackage[inkscapeformat=png]{svg}
\usetikzlibrary{positioning, arrows.meta, decorations.markings}

\input{styles.tex}

\journal{Computers in Biology and Medicine}

\begin{document}
\begin{frontmatter}
\title{Scaling to Multimodal and Multichannel Heart Sound Classification with Synthetic and Augmented Biosignals}

\author[inst1]{Milan Marocchi\corref{cor1}}
\cortext[cor1]{Corresponding author: milan.marocchi@postgrad.curtin.edu.au}
\author[inst1]{Matthew Fynn}
\author[inst2]{Kayapanda Mandana}
\author[inst1]{Yue Rong}
\address[inst1]{School of Electrical Engineering, Computing, and Mathematical Sciences (EECMS), 
Faculty of Science and Engineering, Curtin University, Bentley, WA 6102, Australia}
\affiliation[inst2]{organization={Department of Cardiology, Fortis Healthcare}, city={Kolkata}, postcode={7007107}, state={West Bengal}, country={India}}

\begin{abstract}
\noindent Cardiovascular diseases (CVDs) are the leading cause of death worldwide, accounting for approximately 17.9 million deaths each year. Early detection is critical, creating a demand for accurate and inexpensive pre-screening methods. Deep learning has recently been applied to classify abnormal heart sounds indicative of CVDs using synchronised phonocardiogram (PCG) and electrocardiogram (ECG) signals, as well as multichannel PCG (mPCG). However, state-of-the-art architectures remain underutilised due to the limited availability of synchronised and multichannel datasets.
Augmented datasets and pre-trained models provide a pathway to overcome these limitations, enabling transformer-based architectures to be trained effectively on smaller datasets. 

\noindent This work combines traditional signal processing with denoising diffusion models, WaveGrad and DiffWave, to create an augmented dataset to fine-tune a Wav2Vec 2.0-based classifier on multimodal and multichannel heart sound datasets.

\noindent The approach achieves state-of-the-art performance. On the Computing in Cardiology (CinC) 2016 dataset of single channel PCG, accuracy, unweighted average recall (UAR), sensitivity, specificity and Matthew's correlation coefficient (MCC) reach 92.48\%, 93.05\%, 93.63\%, 92.48\%, and 0.8283, respectively.
Using the synchronised PCG and ECG signals of the training-a dataset from CinC, 93.14\%, 92.21\%, 94.35\%, 90.10\%, and 0.8380 are achieved for accuracy, UAR, sensitivity, specificity and MCC, respectively. Using a wearable vest dataset consisting of mPCG data, the model achieves 77.13\% accuracy, 74.25\% UAR, 86.47\% sensitivity, 62.04\% specificity, and 0.5082 MCC. 

\noindent These results demonstrate the effectiveness of transformer-based models for CVD detection when supported by augmented datasets, highlighting their potential to advance multimodal and multichannel heart sound classification.
\end{abstract}



\end{frontmatter}

\section{Introduction}

Cardiovascular diseases (CVDs) are the leading cause of death globally, accounting for approximately 17.9 million deaths each year \cite{who}.
Since CVD treatment is most effective when the condition is detected early, there is a pressing need for accurate and affordable pre-screening methods.
Cardiac auscultation is one such technique: it is inexpensive, non-invasive, and widely used, relying on physicians listening to heart sounds to detect abnormalities indicative of CVD.
However, auscultation yields relatively low diagnostic accuracy, partly because heart sounds often lie near the threshold of human hearing \cite{heart1, heart2, heart3}.

Recent advances in computer-aided methods have demonstrated promise in improving the accuracy of abnormal heart sound classification.
Combining phonocardiogram (PCG) and electrocardiogram (ECG) signals within deep learning frameworks has produced favourable results \cite{milan}.
However, current datasets suffer from limitations, including class imbalance, low signal-to-noise ratio, and limited size, all hindering robust and accurate classification.
Current state-of-the-art (SOTA) approaches often use convolutional neural networks (CNNs), many of which leverage pre-trained image-based architectures with spectrogram inputs.
Transformer-based models remain underexplored in this domain, with only Vision Transformers (ViTs) evaluated on spectrogram inputs.
Therefore, there is significant potential in applying transformers directly to raw audio signals.

Using raw signals as input offers the advantage of preserving phase information and enabling models to learn features that may be difficult to extract from time-frequency representations.
However, learning time-frequency features from raw audio alone can be challenging.
Modern architectures such as transformers generally require substantial data to outperform existing approaches~\cite{attention}.
Transformer-based models that operate on raw signals show potential for improved performance, but their data requirements must be addressed.

Recent advances in synthetic audio generation using diffusion models offer a promising solution to this data scarcity, enabling easier training of data-hungry models and potentially yielding SOTA performance \cite{diffwave, wavegrad}.
Traditional data augmentation techniques have also been applied to mitigate issues with limited and imbalanced datasets \cite{augmentation}.
Furthermore, fine-tuning large pre-trained models provides another avenue for overcoming data limitations \cite{milan, goutham}.
Models like Wav2Vec 2.0 (Wav2Vec2) \cite{wav2vec} have shown excellent performance in various speech classification tasks \cite{dysarthria}, making them strong candidates for transfer learning in PCG and ECG classification.

A recent development includes a wearable device equipped with up to seven PCG sensors, enabling the collection of synchronised multichannel PCG data \cite{RongFynnNordholm2023PreScreeningCAD}.
This advancement makes it feasible to train models utilising multimodal and multichannel data for abnormal heart sound detection.

This work investigates the use of traditional augmentation, synthetic signal generation, and fine-tuning of large pre-trained models to overcome the data limitations faced by transformer-based approaches.
Specifically, this study fine-tunes Wav2Vec2 for scalable classification, progressing from single-channel PCG signals to synchronised PCG-ECG and multichannel PCG signals.

The paper is structured as follows: Section~\ref{sec:background} provides an overview of the signals and models used.
Section~\ref{sec:materials} describes the datasets and materials.
Section~\ref{sec:methods} details the preprocessing steps, augmentation strategies, signal generation, and model training for each model type.
Section~\ref{sec:resultsdiscussion} presents the results and discusses the performance between each model type and the literature, and concluding remarks in Section~\ref{sec:conclusion}.

The novel contributions are as follows:
\begin{itemize}
    \item  Proposed a scalable transformer-based architecture that supports any number of PCG channels and ECG inputs.
    \item Developed a multichannel PCG diffusion model for synthetic signal generation.
    \item Introduced augmentation techniques tailored for multichannel PCG data.
    \item Achieved state-of-the-art performance on the CinC 2016 training-a dataset, the full CinC 2016 dataset and near-SOTA results on the multichannel vest dataset.
\end{itemize}

\section*{Background}
\label{sec:background}
\subsection{Phonocardiogram and Electrocardiogram Signals}
PCG signals consist of multiple sounds created by sudden changes in blood flow within the heart, causing vibrations \cite{leatham}.
The fundamental heart sounds are S1 and S2.
S1 occurs due to isovolumetric ventricular contraction at the beginning of systole, and S2 results from the closure of the aortic and pulmonic valves at the beginning of diastole.

While S1 and S2 are the most audible sounds, other sounds can also be heard, such as the third heart sound (S3), fourth heart sound (S4), systolic ejection click, mid-systolic click, opening snap, and heart murmurs, which result from turbulent, fast-flowing blood \cite{cinc}.
These sounds all lie in the low-frequency range: S1 ranges from 10–140Hz with the highest energy around 25–45Hz; S2 spans 10–200Hz with energy concentrated around 55–75Hz; and S3 and S4 fall between 20–70Hz, though they are less audible.
Murmurs, which may indicate CVDs, can be found in a wider frequency range from 25Hz to 400Hz \cite{schmidt}, with some extending up to 600Hz but with less energy \cite{springer}.

ECG signals represent the electrical activity of the heart \cite{heart4}.
An ECG signal comprises P waves, QRS complexes, and T waves, with a U wave occasionally present \cite{heart5}.
These components contain diagnostic information useful in identifying CVDs.
ECG signals are typically filtered between 0.5Hz and 40Hz to remove baseline wander and unwanted noise \cite{ecgfiltering}.
In patients with coronary artery disease (CAD), symptoms such as T-wave inversion, ST-T abnormalities, left ventricular hypertrophy, and premature ventricular contractions have been documented \cite{ecgcad}.

\begin{figure}[H]
\centering
\includegraphics[width=1\linewidth]{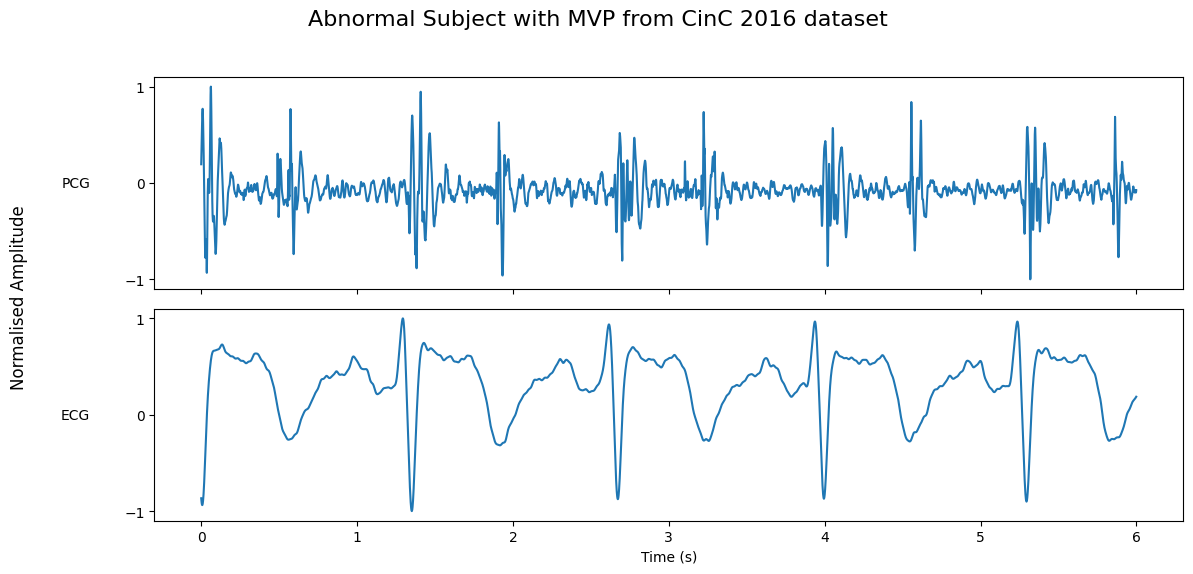}
\caption{PCG and ECG of a patient with an abnormal heart condition.}
\label{fig:patient}
\end{figure}

Combining PCG and ECG signals has yielded better results than using either signal alone \cite{milan}, as both contain complementary diagnostic features.
Figure~\ref{fig:patient} illustrates these signals in a patient with mitral valve prolapse.

\begin{figure}[H]
\centering
\includegraphics[width=1\linewidth]{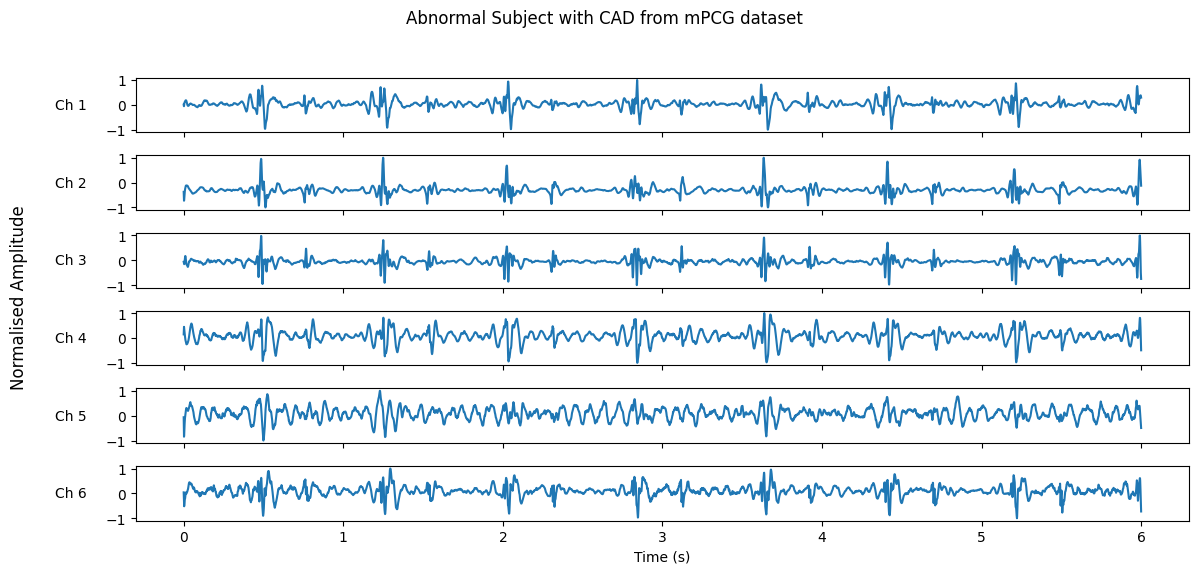}
\caption{mPCG of a patient with an abnormal heart condition.}
\label{fig:patient2}
\end{figure}

Combining multiple PCG sensors in different auscultation regions can yield better results than single-channel PCG, as each channel offers a different resolution of the region where the murmur originates~\cite{vestpaper}. Figure~\ref{fig:patient2} shows a CAD patient with multichannel PCG (mPCG) data.

\subsection{Wav2Vec 2.0}
Wav2Vec 2.0 (Wav2Vec2), developed by Meta, is a speech-to-text model with a transformer-based architecture.
It consists of three primary components: a feature encoder, transformer encoder and a quantisation module for pre-training \cite{wav2vec}.

The feature encoder is a CNN comprising seven convolutional blocks with 512 channels and various strides and kernel sizes, achieving a 20ms stride and a receptive field covering 400 input samples (25ms).
The architecture is available in two variants: BASE and LARGE.
The BASE variant includes 12 transformer blocks with a model dimension of 768, a multilayer perceptron (MLP) dimension of 3072, and 8 attention heads.
The LARGE variant doubles the transformer layers to 24, with a model dimension of 1024, MLP dimension of 4096, and 16 attention heads, enhancing capacity for complex tasks.

The quantisation module discretises and then encodes the output using a Gumbel-Softmax function during self-supervised training.
Training proceeds in two phases: first, self-supervised training on large unlabelled audio datasets (e.g., LibriSpeech \cite{librispeech}) to learn speech representations; second, fine-tuning with 960 hours of labelled speech data across diverse accents and languages \cite{wav2vec}.

\begin{figure}[H]
\centering
\begin{subfigure}{.3\textwidth}
\centering
\includegraphics[width=.4\linewidth]{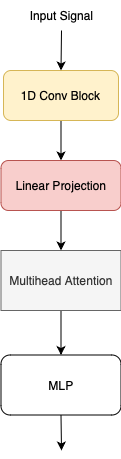}
\caption{Wav2Vec 2.0 encoder architecture.}
\label{fig:dblock}
\end{subfigure}%
\begin{subfigure}{.3\textwidth}
\centering
\includegraphics[width=.4\linewidth]{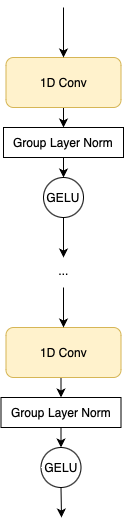}
\caption{CNN module architecture.}
\label{fig:cnnblock}
\end{subfigure}%
\begin{subfigure}{.3\textwidth}
\centering
\includegraphics[width=.25\linewidth]{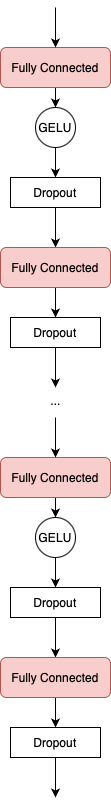}
\caption{MLP module architecture.}
\label{fig:mlp}
\end{subfigure}%
\caption{Wav2Vec 2.0 encoder module architectures.}
\label{fig:wav2vec2modules}
\end{figure}

Pretrained on speech, Wav2Vec2 can be fine-tuned for downstream tasks such as classifying abnormal heart sounds by using its encoder as a feature extractor. This encoder is shown with its main components in Figure~\ref{fig:wav2vec2modules}. The BASE variant of the feature-encoder and transformer encoder is used throughout this work, referred to as the Wav2Vec 2.0 encoder.

\subsection{Diffusion Models}
Diffusion probabilistic models are generative models that convert a simple distribution (e.g., isotropic Gaussian) into a complex data distribution via a Markov chain process \cite{sohldickstein, ho2020}.
Training is based on optimising the variational lower bound (ELBO), even when the data likelihood is intractable.

These models, applied in audio and image synthesis, relate to denoising score matching and do not require separate encoder or discriminator networks like VAEs \cite{kingma2014} or GANs \cite{goodfellow2014}.
This advantage avoids issues such as posterior collapse and mode collapse, making diffusion models particularly effective for high-fidelity audio generation by whitening training data latents with a parameter-free noise process.

\subsubsection{WaveGrad}
\begin{figure}[H]
\centering
\includegraphics[width=0.45\linewidth]{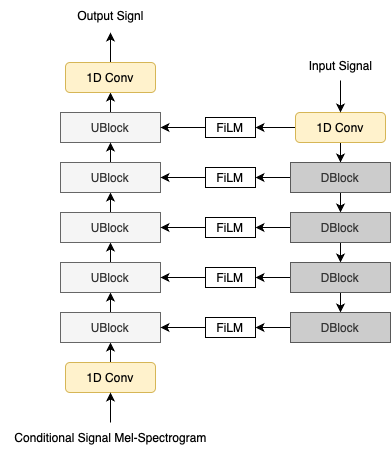}
\caption{WaveGrad architecture.}
\label{fig:wavegrad}
\end{figure}

\begin{figure}[H]
\centering
\begin{subfigure}{.3\textwidth}
\centering
\includegraphics[width=\linewidth]{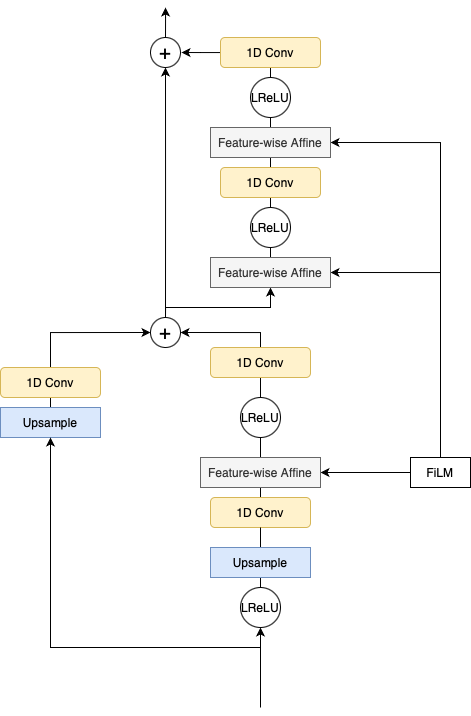}
\caption{UBlock module architecture.}
\label{fig:ublock}
\end{subfigure}%
\begin{subfigure}{.3\textwidth}
\centering
\includegraphics[width=.8\linewidth]{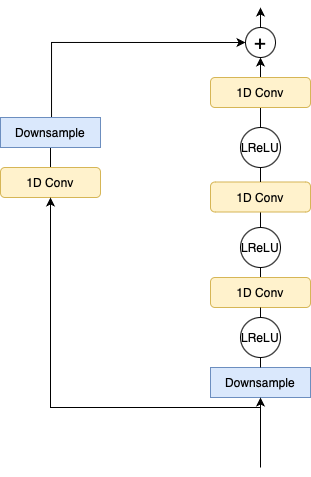}
\caption{DBlock module architecture.}
\label{fig:dblock}
\end{subfigure}%
\begin{subfigure}{.3\textwidth}
\centering
\includegraphics[width=.7\linewidth]{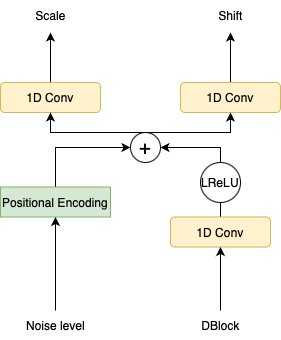}
\caption{FiLM module architecture.}
\label{fig:film}
\end{subfigure}%
\caption{WaveGrad module architectures.}
\label{fig:wavegradmodules}
\end{figure}

WaveGrad is a diffusion model for conditional audio synthesis \cite{wavegrad}.
It employs upsampling (UBlocks) and downsampling (DBlocks) blocks, conditioned on mel-spectrogram inputs, along with Feature-wise linear modulation (FiLM) modules.
These blocks resemble those in the GAN-TTS model \cite{gan-tts}.
Figure~\ref{fig:wavegrad} shows the overall architecture. Module components are shown in Figure~\ref{fig:wavegradmodules}.
The loss function measures the difference between the added noise in the forward diffusion and the predicted noise during denoising.

\subsubsection{DiffWave}
DiffWave is a diffusion model for raw audio synthesis with both conditional and unconditional variants \cite{diffwave}.
The model uses 1D convolutions and fully connected layers, with its core comprising bi-directional dilated convolutions and residual connections.
It is trained using a single ELBO-based objective without auxiliary losses.
Conditional generation uses local conditioning signals and global conditioning via discrete labels.

\begin{figure}[H]
\centering
\includegraphics[width=0.7\linewidth]{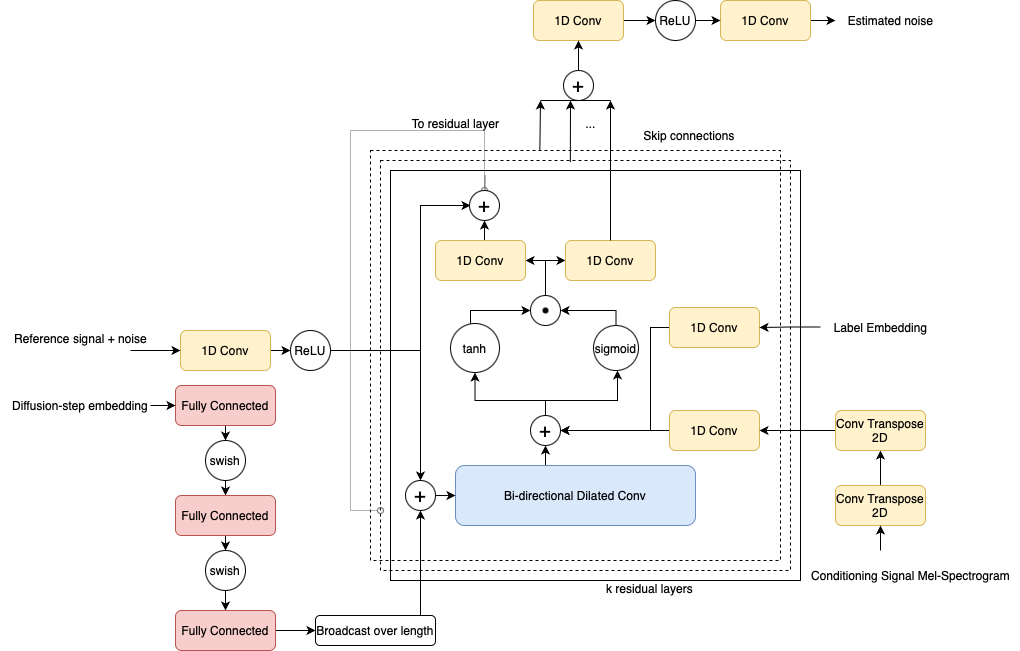}
\caption{DiffWave architecture.}
\label{fig:diffwave}
\end{figure}

\section{Materials}
\label{sec:materials}
All data processing and model training were conducted using a Ryzen 7 3800X CPU and an Nvidia RTX 3090 (24~GB), with Python 3.10 and PyTorch 2.1.2.
Diffusion models were trained on an RTX 4090 using the vast.ai cloud service.

\subsection{Dataset}

\subsubsection{Multimodal dataset}
The classification dataset is sourced from the 2016 PhysioNet Computing in Cardiology Challenge (CinC), comprising five databases (training-a to training-e) \cite{cinc}.
There are 3153 recordings sourced from 764 patients \cite{cinc}, some of which are very noisy. Of these recordings, 665 are abnormal and 2488 are normal, with each recording lasting 5–120 seconds.
Database training-a includes synchronised ECG and PCG recordings; out of 409 recordings, 405 contain both signal types (288 abnormal, 117 normal, hence, we conduct analysis on PCG signal only models using the entire dataset, and PCG-ECG hybrid models using only training-a.

\begin{table}[H]
\centering
\caption{Datasets from CinC 2016 (adapted from \cite{cinc})}
\label{table:datasets}
\begin{tabular}{lcccc}
\toprule
Database & Source Data & Abnormal (\%) & Normal (\%) & Unsure (\%) \\
\midrule
training-a     & MITHSDB     & 67.5 & 28.4 & 4.2 \\
training-b     & AADHHSDB    & 14.9 & 60.2 & 24.9 \\
training-c     & AUTHHSDB    & 64.5 & 22.6 & 12.9 \\
training-d     & UHAHSDB     & 47.3 & 47.3 & 5.5 \\
training-e     & DLUTHSDB    & 7.1  & 86.7 & 6.2 \\
training-f     & SUAHSDB     & 27.2 & 68.4 & 4.4 \\
\textbf{All training} &              & 18.1 & 73.0 & 8.8 \\
\bottomrule
\end{tabular}
\end{table}

A 60-20-20 split is used for training, validation, and testing in both the combined and training-a subsets.
Synthetic data is generated from training-a and used only in models trained on this subset due to the need for ECG-conditioned signals.

\subsubsection{Multichannel dataset}
Recordings were obtained from subjects using a multichannel wearable vest with seven phonocardiogram sensors recording at different auscultation sites \cite{vestpaper}. Ninety-six subjects were diagnosed with coronary artery disease through angiography, and an additional 61 were subjects without the disease. Of the 61, 21 were control subjects below the age of 35, assumed to be normal, as the risk of CAD is significantly higher in people aged 45 and over. The multichannel device had sixty seconds of data recorded in a hospital with background noise, with normal breathing occuring. This vest dataset also did not have optimal positioning of all the stethoscopes, resulting in lower signal-to-noise ratio (SNR) as compared to the other datasets in this study, making this dataset representative of real-world data. This work only utilises the front six channels, as the back channel was found to be contaminated with breathing noise.

As this dataset contains less data than the other two datasets, a seven-fold cross-validation was used. For each iteration, a different fold is the validation set, and another is the test set, with all the others used for training. No two folds are used for test or validation twice. The folds are stratified to ensure each fold contains the same proportion of CAD to normal subjects.

\subsubsection{Generative and augmentation datasets}
In addition to training-a, the Icentia dataset \cite{icentia} was used to provide novel ECG inputs for generating PCG signals.
This dataset contains 11,000 patients and 2.77 billion labelled heartbeats sampled at 250~Hz, with 541,794 segments.
Each beat is labelled as normal, premature atrial contraction, premature ventricular contraction, or one of several rhythm types (sinus, atrial fibrillation, atrial flutter).

To enhance robustness to noise, augmentations were performed using additional datasets: EPHNOGRAM (for PCG) and the MIT-BIH Noise Stress Test Database (for ECG).
EPHNOGRAM comprises PCG recordings from 24 healthy adults during rest and stress conditions \cite{ephnogram}.
The MIT-BIH dataset contains 12 half-hour ECG recordings and three half-hour recordings of typical noise, including baseline wander, muscle artefacts, and electrode motion \cite{mit}.
These noise samples were used to augment ECG signals.

\section{Methods}
\label{sec:methods}
\begin{figure}[htbp]
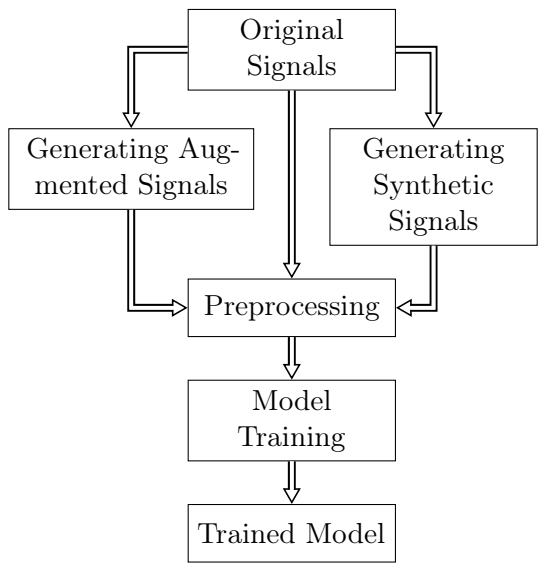

\centering
\subfile{Figures/ModelProcedure.tex}
\caption{Classification model creation procedure.}
\label{fig:method}
\end{figure}

This study aimed to establish a baseline for both single-input, multimodal and multichannel classification models, evaluate the effectiveness of traditional and synthetic data augmentation, and assess the scalability of the approach to larger datasets.
A step-by-step procedure including signal augmentation, synthetic signal generation, preprocessing, and classification was followed as shown in Figure~\ref{fig:method}.

\subsection{Augmented Signals}

\subsubsection{Single channel augmentations}
\begin{figure}[htbp]
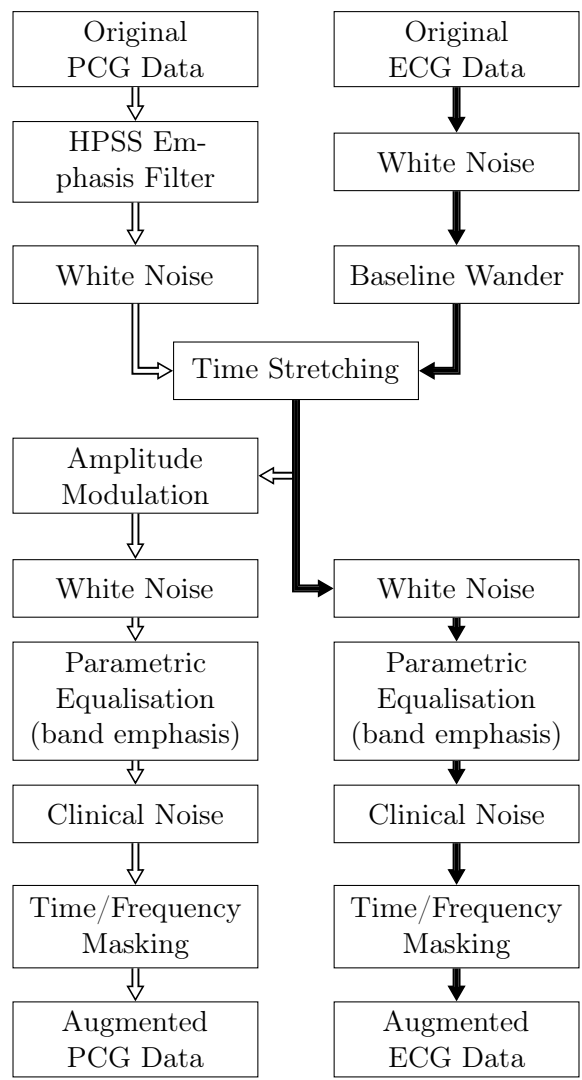

\centering
\subfile{Figures/EPCGAugmentation.tex}
\caption{PCG and ECG augmentation procedure}
\label{fig:augmentation}
\end{figure}

The PCG and ECG augmentation process is shown in Figure~\ref{fig:augmentation} (adapted from \cite{epcgdiffusion}).
Each augmentation has a specific probability of occurring per sample: harmonic-percussive source separation (75\%), white noise (7.5\%), time stretching (25\%), amplitude modulation (75\%), baseline wander (75\%), parametric equalisation (25\%), and clinical noise (50\%). 

Two additional online augmentations were applied with 20\% probability during training: (1) time and frequency masking, and (2) additional time stretching. These augmentations aid in regularisation and generalisation.

\subsubsection{Multichannel augmentations}
\begin{figure}[htbp]
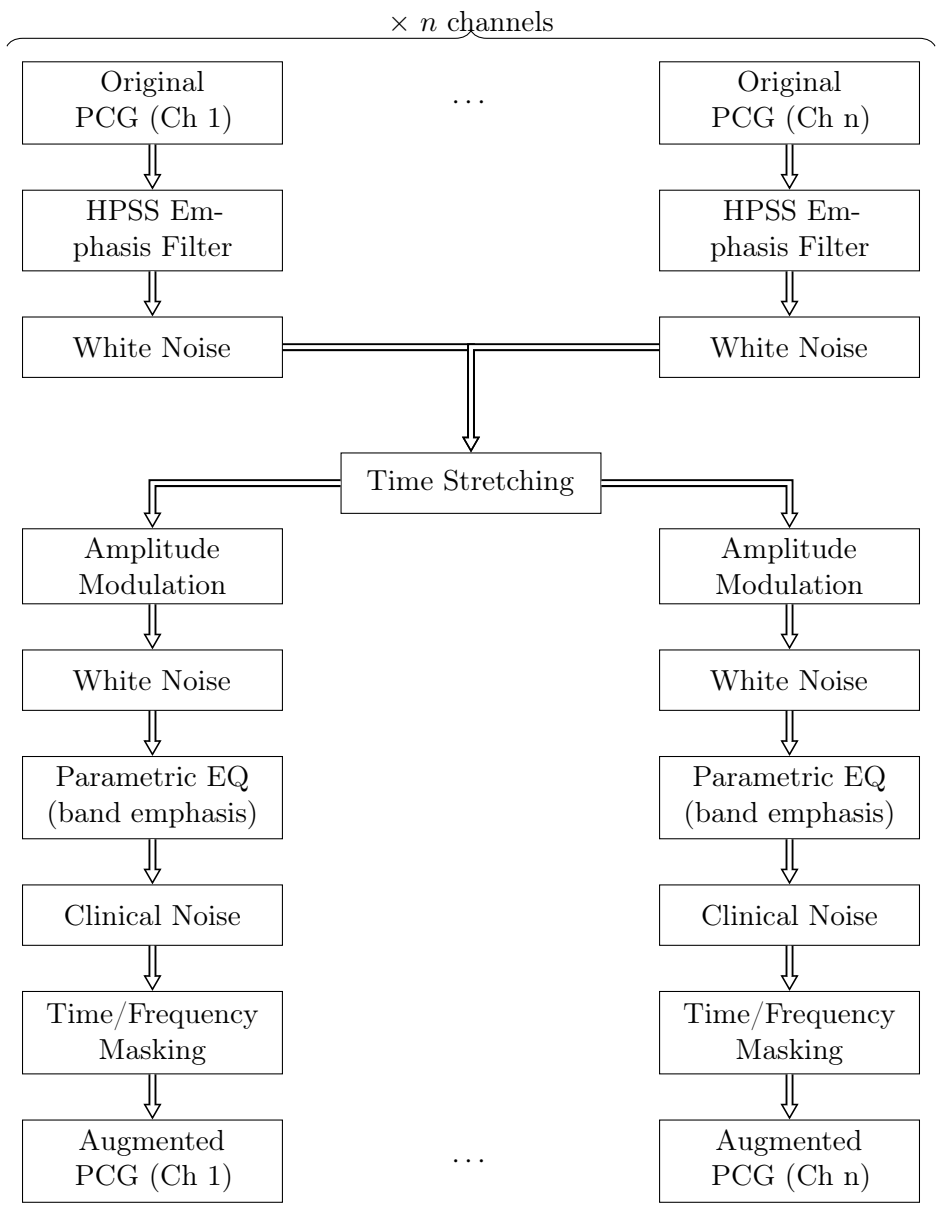

\centering
\subfile{Figures/mPCGAugmentation.tex}
\caption{Multichannel PCG augmentation procedure}
\label{fig:maugmentation}
\end{figure}

The mPCG followed similar augmentations to the single channel augmentations \cite{epcgdiffusion}, but was further extended from \cite{epcgdiffusion} to support mPCG signals, by synchronising the time-stretching across channels the same way as it was for synchronising the PCG and ECG data. All other augmentations were applied independently to each channel. Figure~\ref{fig:maugmentation} shows the procedure for multichannel PCG data augmentation, with the time-stretching augmentation being synchronised across all channels in the augmented patient.

\subsection{Synthetic Signals}

\subsubsection{Single channel PCG}
\begin{figure}[htbp]
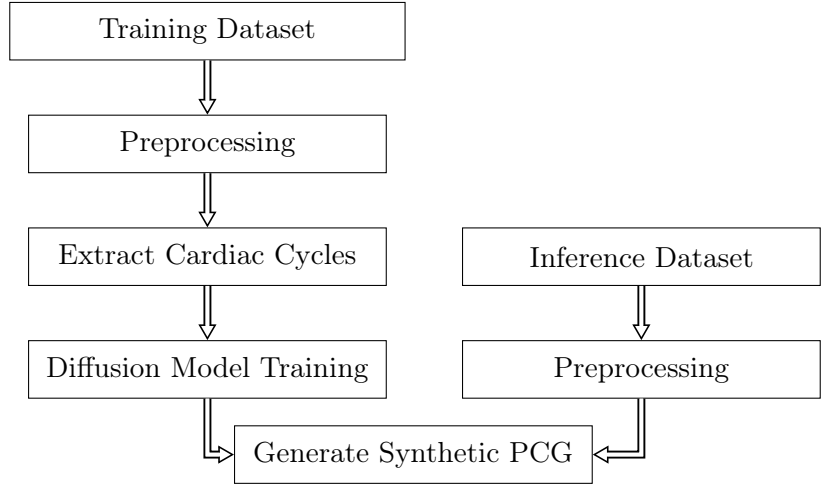

\centering
\subfile{Figures/SyntheticProcedure.tex}
\caption{Synthetic signal generation procedure.}
\label{fig:synthetic}
\end{figure}

Synthetic PCG signals were generated using ECG signals from the Icentia dataset \cite{icentia} as conditioners, following \cite{epcgdiffusion}.
WaveGrad \cite{wavegrad} and DiffWave \cite{diffwave} generated 3,200 patients' data each, using a 3:1 ratio of normal to abnormal samples.
Local conditioning was done using ECG mel-spectrograms, and global conditioning was done using disease labels.
The generation process is summarised in Figure~\ref{fig:synthetic}.

Training lasted 24 hours on an RTX 3090, with cardiac cycle rearrangement (25\% probability) applied during training to reduce overfitting. Rearrangement types include shuffling large groups, 1–4 heart cycle chunks, and individual cycles \cite{epcgdiffusion}. Crossfading was used to minimise artifacts.

Signals were bandpass filtered (2–500Hz for PCG, 0.25–100Hz for ECG) and resampled to 4~kHz.
Mel-spectrograms for ECG were computed with a 1024 window length, 256 hop length, and 80 mel bins.

\subsubsection{Multichannel PCG}
The same diffusion models were utilised to generate multichannel PCG, also following the same procedure from Figure~\ref{fig:synthetic}, however, the global conditioning labels were modified in order to go from any channel as the local conditioning channel to any other output channel. Hence, the label would specify the conditioning channel and the reference/generated channel alongside the subject's condition; normal or CAD.

Other than this modification, the process for training the multichannel diffusion models remained the same.

Synthetic mPCG signals were generated from conditioning PCG data from the CinC 2016 dataset, utilising training-a and training-b to generate new multichannel subjects. These were used as they contained information on their ausculation sites as well as training-b, being a CAD database, so it could be used to generate synthetic CAD subjects.

\subsection{Preprocessing and Segmentation}

Signals were resampled from 2kHz to 1kHz, bandpass filtered (25–400Hz for PCG, 2–60Hz for ECG), min-max normalised, then resampled to 4.125kHz or 16kHz for classification.
A grid search from 1–16kHz (in 125Hz increments) identified 4.125kHz as optimal for the training-a and vest datatsets, but 16kHz for the CinC 2016 dataset.

For the training-a and the CinC dataset signals were segmented into 4-second overlapping windows, whereas the vest dataset had 2-second overlapping windows. The windows for all datasets consisted of 0.25s overlap. The first 0.3s of each recording was excluded to remove artifacts found at the start of signals.
In the training-a and the CinC dataset synthetic signals were limited to two segments per recording to reduce overfitting, randomly chosen from the signals.
The vest dataset used fewer synthetic signals, due to limited PCG signals databases with annotated auscultation positions, so all segments were utilised.
Final classification was based on averaging fragment-level predictions.

\subsection{Models}
Three model types were evaluated: single-input (PCG or ECG), multimodal (PCG+ECG) and multichannel (multichannel PCG). All used the Wav2Vec2 BASE encoder to extract 768 features per input. They all use the same Wav2Vec2 feature extractors, with the multimodal and multichannel model concatenating the extracted features from each representation. All models are very similar, hence, can be scaled from single channel to multimodal and multichannel.

From training, the chosen model is based on the model's performance on the validation set, with the highest Matthew's correlation coefficient (MCC) value.
This value is used as it combines much information about how well the model performs into a single value~\cite{mcc}. While not fully comprehensive, it resulted in the best-performing models when used as the main selection metric.
The MCC, as well as other metrics, are shown below and is used to evaluate model performance against the test set, as the MCC value does not fully summarise the confusion matrix.
TP represents true positives, TN true negatives, FP false positives, and FN false negatives.

\begin{equation}
\text{Sensitivity (TPR)} = \frac{TP}{TP+FN}
\end{equation}

\begin{equation}
\text{Specificity (TNR)} = \frac{TN}{TN+FP}
\end{equation}

\begin{equation}
\text{False Positive Rate (FPR)} = \frac{FP}{TP+FP}
\end{equation}

\begin{equation}
\text{Accuracy (acc)} = \frac{TP+TN}{TP+TN+FP+FN}
\end{equation}

\begin{equation}
\text{Unweighted average recall (UAR)} = \frac{TPR+TNR}{2}
\end{equation}

\begin{equation}
\text{F1 positive score (F1)} = \frac{2 \times TP}{2 \times TP + FP + FN}
\end{equation}

\begin{equation}
\text{MCC} = \frac{TP \cdot TN - FP \cdot FN}{\sqrt{(TP+FP)(TP+FN)(TN+FP)(TN+FN)}}
\end{equation}

Table~\ref{table:models} details all of the models that are to be trained and evaluated to demonstrate this scalable architecture. This includes single-input models, multimodal models and multichannel models.

\begin{table}[H]
\footnotesize
\centering
\captionsetup{skip=10pt}
\caption{Models to be trained and evaluated.}
\label{table:models}
\begin{tabular}{l l l l}
\toprule
\textbf{Dataset} & \textbf{Inputs} & \textbf{Sampling rate} & \textbf{Data} \\
\midrule
training-a & PCG+ECG & 16kHz & Original \\
training-a & PCG+ECG & 16kHz & All \\
training-a & PCG+ECG & 4.125kHz & All \\
CinC & PCG & 16kHz & Original \\
CinC & PCG & 16kHz & All \\
CinC & PCG & 4.125kHz & All \\
Vest Data & mPCG & 16kHz & Original \\
Vest Data & mPCG & 4.125kHz & All \\
Vest Data & mPCG & 16kHz & All \\
\bottomrule
\end{tabular}
\end{table}

Hyperparameters were tuned using a Bayesian optimisation (Optuna~\cite{optuna}) on the augmented datasets. The average (across five runs, to account for the variation of training the models) MCC score on the validation set was utilised as the optimisation metric. 
The initial hyperparameters for the baseline model without the augmented dataset are found in Table~\ref{table:base_params_s}.
This optimisation used 10 epochs from the original data with at least 30 augmentations per patient for the single channel and multimodal models, following the first step of the training schedule that the models undergo.

There were 150 trials (each run 5 times) run within the optimisation, with the main objective being the highest average MCC of the model in the validation set.
For the multichannel models, the average MCC value also included averaging over all seven folds.
It was also found that the hyperparameters generalised to other shuffled train-validation-test splits.
The number of neurons per layer in the fully connected layers was optimised, as well as the hyperparameters for the learning rate scheduler and the optimiser.

The following sections will detail the architecture and training of each type of model, along with the hyperparameters used.

\subsubsection{Single channel models}
\label{sec:singlechannelmodels}
\begin{figure}[H]
\footnotesize
\centering
\includegraphics[width=0.8\linewidth]{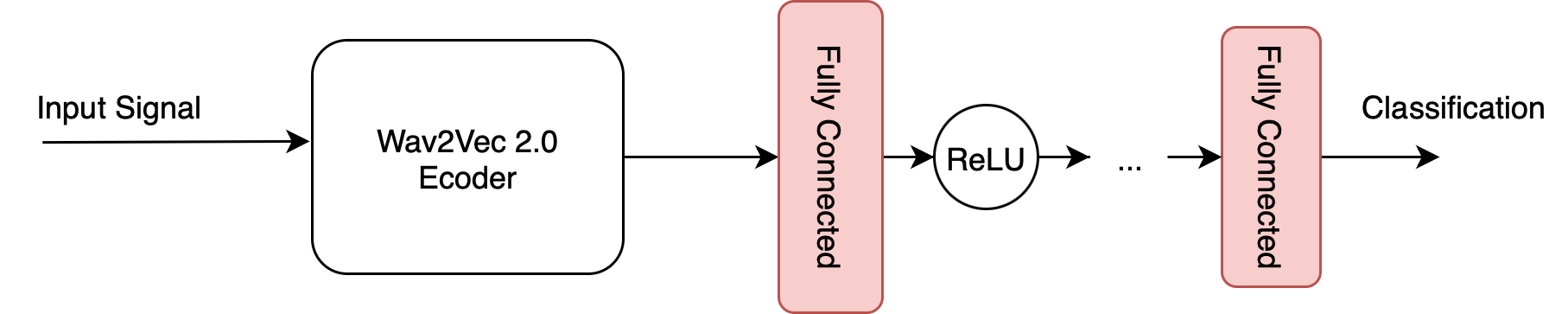}
\caption{Single input model architecture.}
\label{fig:singlearch}
\end{figure}

The architecture of the single-channel model is displayed in Figure~\ref{fig:singlearch}. Training of the single-channel models followed the schedule in Table~\ref{table:schedule}. Initial training was performed on original data, followed by synthetic and augmented data to prevent overfitting. To further reduce overfitting to the synthetic data, only three segments are extracted from each synthetic subject.

\begin{table}[H]
\footnotesize
\centering
\caption{Training schedule for the training-a and the CinC datasets.}
\label{table:schedule}
\begin{tabular}{cccc}
\toprule
\textbf{Data} & \textbf{Epochs} & \textbf{Normal augments} & \textbf{Abnormal augments} \\
\midrule
Original                    & 10 & 60 & 30 \\
DiffWave                   &  4 & 30 &  5 \\
Original                   &  2 & 60 & 30 \\
Original/DiffWave/WaveGrad &  4 & 30/5/5 & 30/5/5 \\
WaveGrad                   &  4 & 30 &  5 \\
Original                   &  2 & 60 & 30 \\
\bottomrule
\end{tabular}
\end{table}

The stochastic gradient descent (SGD) optimiser was used with a step exponential decay learning rate scheduler, with the learning rate, weight decay and batch size being hyperparameters from the optimiser, and the momentum, gamma and step sizes being from the learning rate scheduler. Initial model hyperparameters of the single channel models for baseline and augmented datasets are listed in Tables~\ref{table:base_params_s} and \ref{table:params_s}.

\begin{table}[H]
\footnotesize
\centering
\caption{Baseline single channel model hyperparameters.}
\label{table:base_params_s}
\hspace*{-1cm}
\begin{tabular}{l l l l l l}
\toprule
\textbf{Hyperparameter} &   \textbf{CinC PCG Model}  \\
\midrule
Learning rate & 0.001  \\
Weight decay & 1e-5  \\
Momentum & 0.9  \\
Gamma & 0.1  \\
Step size & 3 \\
Batch size & 64 \\
Number of hidden layers & 1 \\
Hidden layer size & 512 \\
\bottomrule
\end{tabular}
\end{table}

\begin{table}[H]
\footnotesize
\centering
\caption{Augmented dataset single channel model hyperparameters.}
\label{table:params_s}
\hspace*{-1cm}
\begin{tabular}{l l l l l l}
\toprule
\textbf{Hyperparameter} &  \textbf{CinC PCG Model}  \\
\hline
\midrule
Learning rate & 0.001 \\
Weight  decay & 4.11e-5 \\
Momentum  & 0.57562  \\
Gamma & 0.167    \\
Step size & 2 \\
Batch size  & 32 \\
Number of hidden layers  & 3 \\
Hidden layer size & 512 \\
\bottomrule
\end{tabular}
\end{table}

\subsubsection{Multimodal models}

\begin{figure}[H]
\centering
\includegraphics[width=0.8\linewidth]{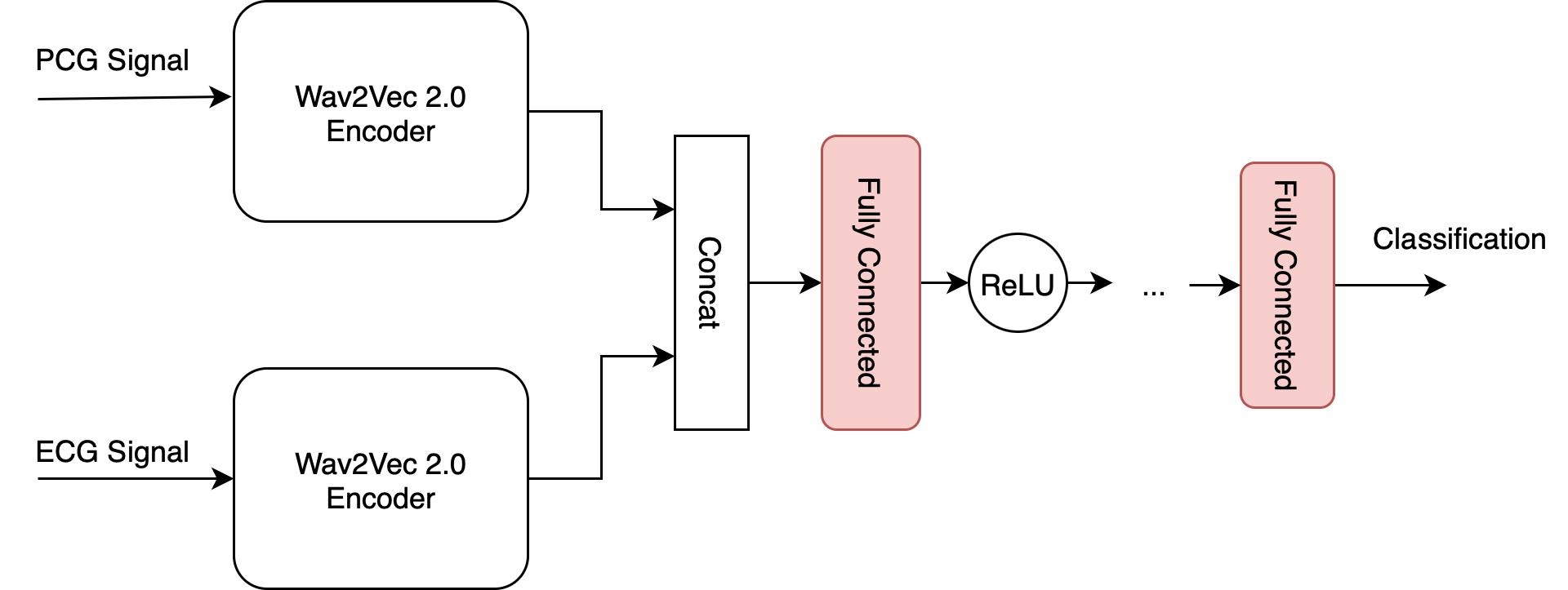}
\caption{PCG and ECG model architecture.}
\label{fig:multiarch}
\end{figure}

The multimodal architecture is found in Figure~\ref{fig:multiarch}. The multimodal models followed the same schedule as the single channel models in Table~\ref{table:schedule}. Firstly, a single channel model is trained on each of the modalities, following the training and architecture in Section~\ref{sec:singlechannelmodels}. These models are trained only using the first dataset from the schedule in Table~\ref{table:schedule}. Following this, the feature extractors from these single-channel models are used for the feature extractor of the multimodal model, with a new MLP classification head attached as in Figure~\ref{fig:multiarch}. The entire model is then trained following the schedule in Table~\ref{table:schedule}. This was done to allow for easier training of the multimodal models, as the feature extractors require fewer updates to their parameters.

This model was trained using the SGD optimiser with an exponential learning rate decay, with the same hyperparameters as the single channel model. The initial hyperparameters of the multimodal models are found in Table~\ref{table:base_params_modal}, with the augmented datasets hyperparameters in Table~\ref{table:params_modal}.

\begin{table}[H]
\footnotesize
\centering
\caption{Baseline multimodal model hyperparameters.}
\label{table:base_params_modal}
\hspace*{-1cm}
\begin{tabular}{l l l l l l}
\toprule
\textbf{Hyperparameter} & \textbf{Multimodal model} \\
\midrule
Learning rate & 0.001 \\
Weight decay & 1e-5 \\
Momentum & 0.9 \\
Gamma & 0.1  \\
Step size & 3 \\
Batch size & 64 \\
Number of hidden layers & 3 \\
Hidden layer size & 512 \\
\bottomrule
\end{tabular}
\end{table}

\begin{table}[H]
\footnotesize
\centering
\caption{Augmented dataset multimodal model hyperparameters.}
\label{table:params_modal}
\hspace*{-1cm}
\begin{tabular}{l l l l l l}
\toprule
\textbf{Hyperparameter} & \textbf{Multimodal model} \\
\hline
\midrule
Learning rate & 0.001 \\
Weight  decay & 3.11e-5 \\
Momentum & 0.17562 \\
Gamma & 0.002444 \\
Step size & 7 \\
Batch size & 64 \\
Number of hidden layers & 3 \\
Hidden layer size & 1024 \\
\bottomrule
\end{tabular}
\end{table}

\subsubsection{Multichannel models}

\begin{figure}[H]
\centering
\includegraphics[width=0.6\linewidth]{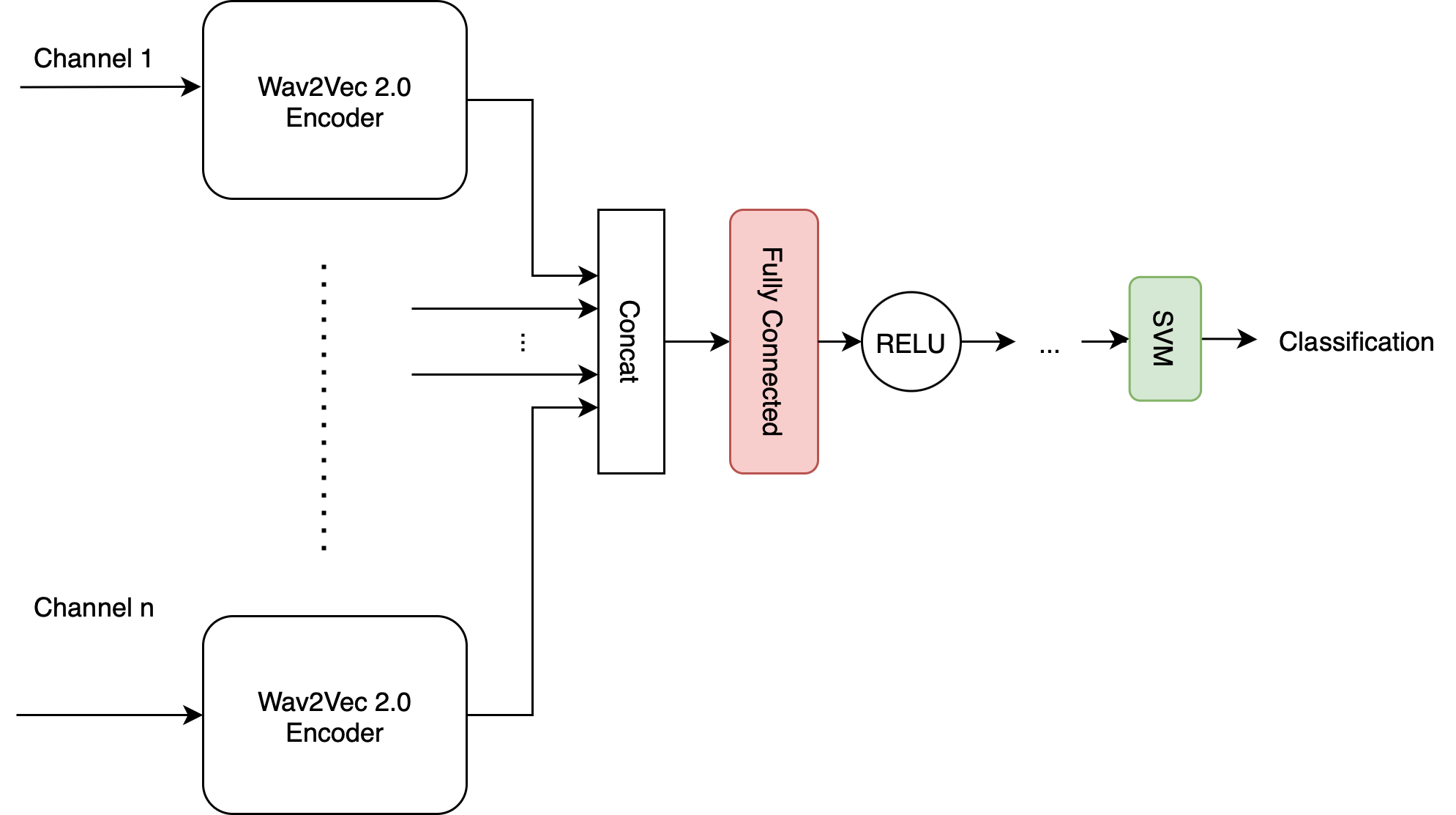}
\caption{Multichannel PCG model architecture.}
\label{fig:multichannelarch}
\end{figure}

The multichannel architecture is shown in Figure~\ref{fig:multichannelarch}.
The multichannel model includes a support vector machine (SVM) that replaces the hidden layers and output layer of the MLP after training has been complete, leaving only the initial fully connected layer of the MLP. Additionally, for fine-tuning, low-rank adaptation (LoRA) \cite{lora} is utilised for the multichannel model due to this dataset containing more limited data than the other datasets. Both the combination of LoRA for fine-tuning and the SVM help to prevent the model from overfitting to the training set.

The multichannel model followed the schedule in Table~\ref{table:schedule2}, where training-a and training-b refer to the synthetic data generated conditioned on the data from those datasets. Only a single synthetic subject was created for each subject in the conditioning dataset. This was done to reduce over-fitting to the synthetic data.

\begin{table}[H]
\footnotesize
\centering
\caption{Training schedule for vest dataset.}
\label{table:schedule2}
\begin{tabular}{cccc}
\toprule
\textbf{Data} & \textbf{Epochs} & \textbf{Normal augments} & \textbf{Abnormal augments} \\
\midrule
Original/training-a                  & 10 & 0 & 0 \\
Orignal/training-a/training-b &  2 & 20/4/4 & 10/2/2 \\
\bottomrule
\end{tabular}
\end{table}

 The training differs as there are now six feature extractors that are to be trained first. Once training has been completed, the outputs from the first layer of the MLP (after the activation function) are taken as inputs to an SVM. This SVM uses the radial basis function kernel with default scikit learn parameters. The neural network is then frozen, and the SVM is fit from the training set. The root mean square propagation (RMSProp) optimiser was used, along with an exponential decay learning rate scheduler for training the multichannel models. The baseline and augmented mPCG dataset hyperparameters are listed in Tables~\ref{table:base_params_multi} and \ref{table:params_multi}.

\begin{table}[H]
\footnotesize
\centering
\caption{Baseline hyperparameters.}
\label{table:base_params_multi}
\hspace*{-1cm}
\begin{tabular}{l l l l l l}
\toprule
\textbf{Hyperparameter} & \textbf{Multichannel model} \\
\midrule
Learning rate &  0.001 \\
Weight decay &  1e-5 \\
Momentum & 0.9  \\
Gamma & 0.1  \\
Step size & 3 \\
Batch size & 64 \\
Number of hidden layers &  1\\
Hidden layer size & 512 \\
\bottomrule
\end{tabular}
\end{table}

\begin{table}[H]
\footnotesize
\centering
\caption{Augmented dataset hyperparameters.}
\label{table:params_multi}
\hspace*{-1cm}
\begin{tabular}{l l l l l l}
\toprule
\textbf{Hyperparameter}  & \textbf{Multichannel model} \\
\hline
\midrule
Learning rate & 1e-5\\
Weight  decay &  6.1148e-05\\
Momentum &  0.17562 \\
Gamma &  0.02444   \\
Step size  & 4\\
Batch size & 32\\
Number of hidden layers & 3\\
Hidden layer size & 512\\
\bottomrule
\end{tabular}
\end{table}

\section{Results \& Discussion}
\label{sec:resultsdiscussion}

This section evaluates the proposed method for scaling a Wav2Vec2 encoder to multiple signals using the augmented dataset. We begin with: (i) a large single-channel dataset to validate baseline performance, scaling to (ii) a smaller and noisier multimodal dataset, and (iii) a real-world multichannel dataset with hospital noise and suboptimal sensor placement. For each model/dataset type, we present the quantitative results, followed immediately by an interpretation of these results, supported by receiver operator characteristic (ROC) curves, pairwise controlled manifold approximation (PaCMAP) embeddings, interpretability plots, and comparisons with the literature.
The ROC curves show the performance of the models (TPR/FPR) as the threshold of the models is adjusted.
PaCMAP embeddings display the embedding space of the model projected to two dimensions, which helps show how well the model has encoded each data point based on their class. Interpretability plots help to show whether relevant features are being utilised for classification, in the case of CVDs, that the appropriate heart cycle phase is being used.

\subsection{Single channel models}

The results for the single channel models trained on the CinC data are presented in Table~\ref{table:cinc_results_fragment} and Table~\ref{table:cinc_results_patient}, for the fragment and subject-level, respectively. Each model was trained and tested ten times over five shuffled train-validation-test splits. The results are reported as mean $\pm$ standard deviation.
The best model for each metric is highlighted.

\begin{table}[H]
\footnotesize
\captionsetup{skip=10pt}
\caption{Model performance at the fragment level on the CinC dataset.}
\label{table:cinc_results_fragment}
\scriptsize
\begin{tabular}{l l l l l l l l l}
\toprule
$f_s$ & \textbf{Data } & \textbf{Acc} & \textbf{UAR} & \textbf{TPR} & \textbf{TNR} & \textbf{F1} & \textbf{MCC} \\
\midrule
16kHz    & Original         & 87.13$\pm$1.64\% & 85.22$\pm$1.47\% & 81.27$\pm$3.24\% & 89.18$\pm$2.55\% & 91.17$\pm$1.25\% & 0.6798$\pm$0.0320 \\
\hline
16kHz    & All              & 90.62$\pm$1.00\% & \textbf{90.67$\pm$0.72\%} & \textbf{90.97$\pm$1.28\%} & 90.42$\pm$1.58\% & 92.92$\pm$0.95\% & \textbf{0.7930$\pm$0.0142} \\
\hline
4.125kHz & All              & \textbf{90.86$\pm$0.41\%} & 90.66$\pm$0.86\% & 90.16$\pm$1.83\% & \textbf{91.11$\pm$0.40\%} & \textbf{93.70$\pm$0.30\%} & 0.7762$\pm$0.1240 \\
\bottomrule
\end{tabular}
\end{table}

\begin{table}[H]
\footnotesize
\captionsetup{skip=10pt}
\caption{Model performance at the subject level on the CinC dataset.}
\label{table:cinc_results_patient}
\scriptsize
\begin{tabular}{l l l l l l l l l}
\toprule
$f_s$ & \textbf{Data } & \textbf{Acc} & \textbf{UAR} & \textbf{TPR} & \textbf{TNR} & \textbf{F1} & \textbf{MCC} \\
\midrule
16kHz    & Original     & 89.33$\pm$1.56\% & 86.42$\pm$1.49\% & 81.28$\pm$3.89\% & 91.55$\pm$2.49\% & 76.65$\pm$2.55\% & 0.7017$\pm$0.0329 \\
\hline
16kHz    & All          & 92.48$\pm$0.97\% & \textbf{93.05$\pm$0.70\%} & \textbf{93.63$\pm$1.87\%} & 92.48$\pm$0.97\% & 94.93$\pm$0.50\% & \textbf{0.8283$\pm$0.0103} \\
\hline
4.125kHz & All          & \textbf{92.98$\pm$0.75\%} & 92.48$\pm$1.43\% & 91.58$\pm$2.73\% & \textbf{93.35$\pm$0.55\%} & \textbf{95.42$\pm$0.50\%} & 0.8064$\pm$0.0226 \\
\bottomrule
\end{tabular}
\end{table}

 Training with the augmented dataset increases performance relative to the unaugmented baseline, with increases of 3.15\%, 6.63\%, and +0.1266 to accuracy, UAR and MCC at the subject-level, respectively. With all CinC data, 16kHz yields slightly higher subject-level UAR (93.05\%) and MCC (0.8283) than 4.125kHz, although 4.125kHz attains the best fragment-level accuracy/TNR/F1. This indicates sampling-rate tuning is worthwhile per dataset, as it modifies the initial representation passed to the encoder, which can benefit different audio tasks with different frequency bands.

\begin{figure}[H]
  \centering
  \begin{subfigure}[b]{0.45\textwidth}
    \includegraphics[width=\textwidth]{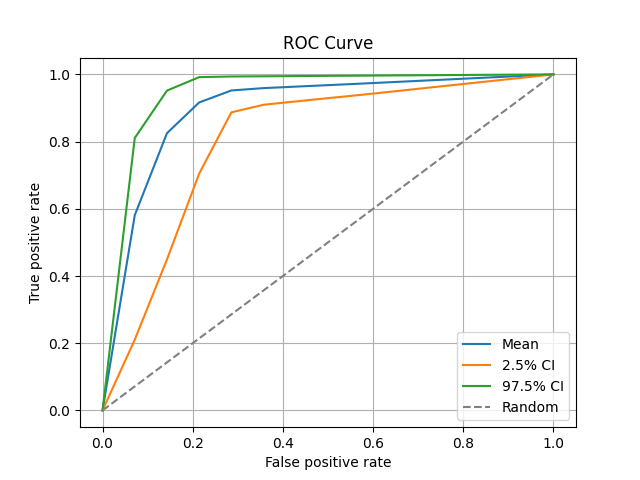}
    \caption{Fragment level.}
  \end{subfigure}
  \hfill
  \begin{subfigure}[b]{0.45\textwidth}
    \includegraphics[width=\textwidth]{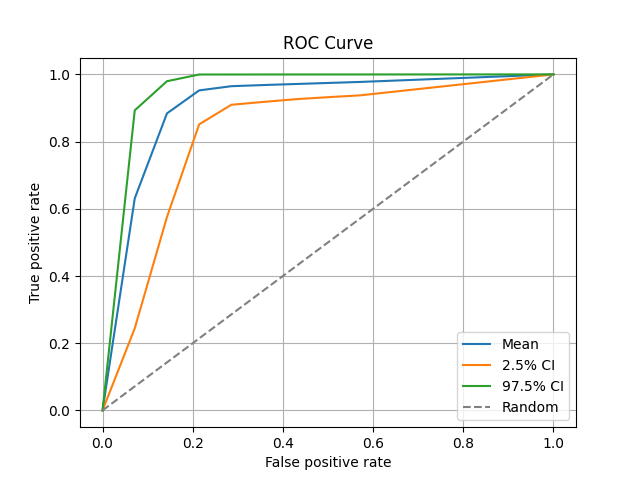}
    \caption{Subject level.}
  \end{subfigure}
  \caption{PCG 16kHz no augments CinC 2016 model ROC plots}
  \label{fig:cinc-noaug-16k}
\end{figure}

\begin{figure}[H]
  \centering
  \begin{subfigure}[b]{0.45\textwidth}
    \includegraphics[width=\textwidth]{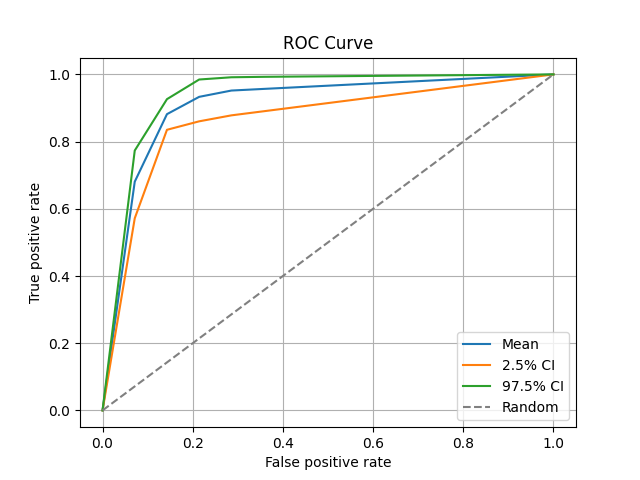}
    \caption{Fragment level.}
  \end{subfigure}
  \hfill
  \begin{subfigure}[b]{0.45\textwidth}
    \includegraphics[width=\textwidth]{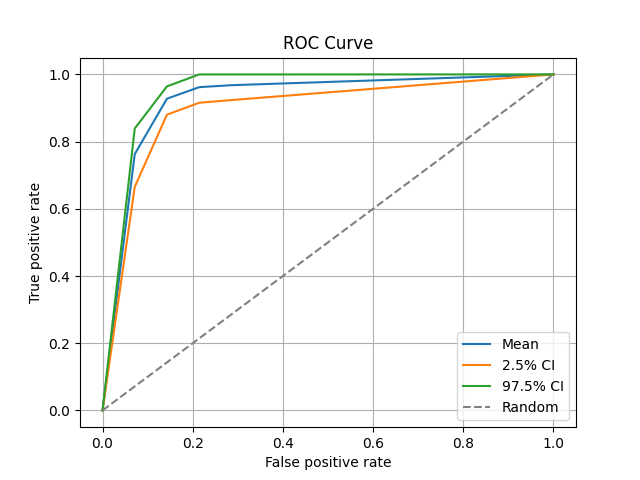}
    \caption{Subject level.}
  \end{subfigure}
  \caption{PCG 16kHz CinC 2016 model ROC plots}
  \label{fig:cinc-16k}
\end{figure}

\begin{figure}[H]
  \centering
  \begin{subfigure}[b]{0.45\textwidth}
    \includegraphics[width=\textwidth]{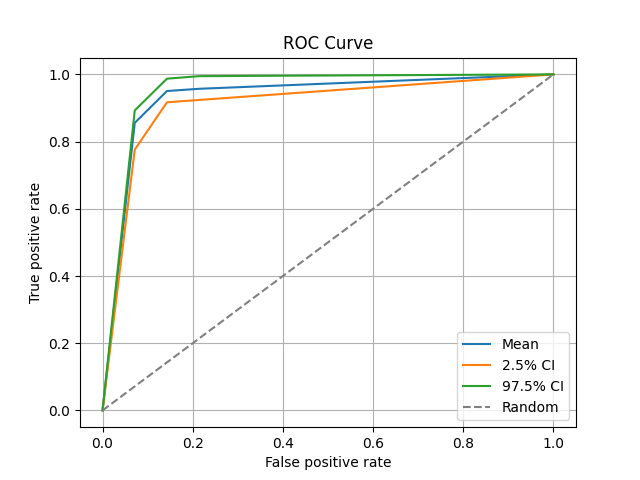 }
    \caption{Fragment level.}
  \end{subfigure}
  \hfill
  \begin{subfigure}[b]{0.45\textwidth}
    \includegraphics[width=\textwidth]{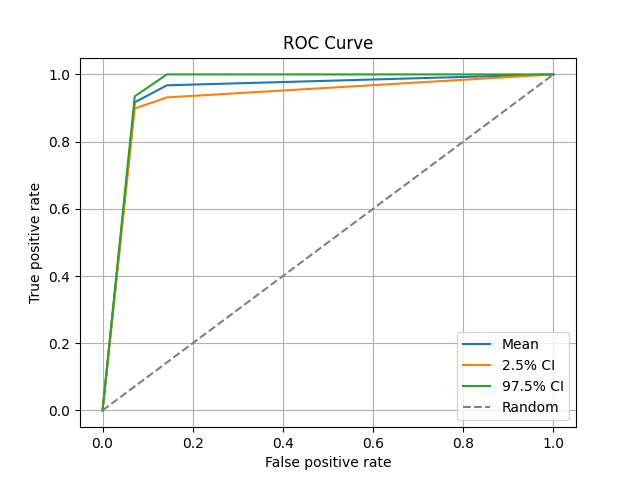}
    \caption{Subject level.}
  \end{subfigure}
  \caption{PCG 4.125kHz CinC 2016 model ROC plots}
  \label{fig:cinc-4125}
\end{figure}

The ROC curves of each of the models are found below, with the mean, 2.5\% confidence interval (CI) and 97.5\% CI being shown to demonstate the variation between models.
For the CinC 2016 models, the ROC curves for; 16kHz with no augments, 16kHz with augments and 4.125kHz with augments, are shown in Figure~\ref{fig:cinc-noaug-16k}, Figure~\ref{fig:cinc-16k} and Figure~\ref{fig:cinc-4125}, respectively.

 Augmentation was shown to yield better ROC curves (less threshold sensitivity) versus the no-augment setting, along with less variation. Additionally, the area under the curve (AUC) is increased with augmentation, with 4.125kHz and 16kHz resulting in similar performance, but 4.125kHz has a slightly higher AUC.

\begin{figure}[H]
    \centering
    \includegraphics[width=0.5\linewidth]{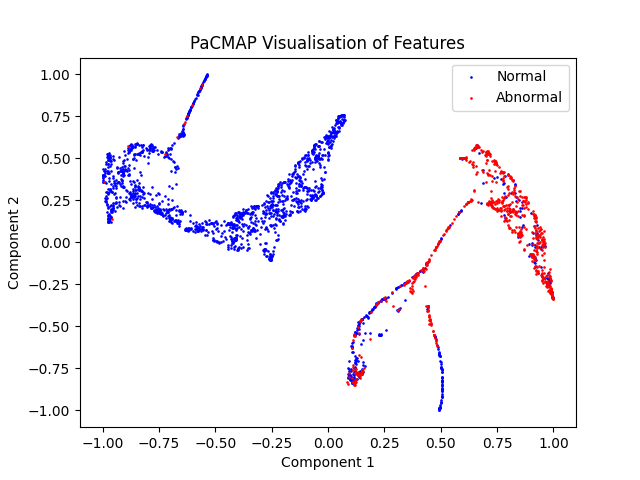}
    \caption{PaCMAP for average no augment 16kHz CinC PCG model.}
    \label{fig:cinc-16k-noaugment-pacmap}
\end{figure}

\begin{figure}[H]
    \centering
    \includegraphics[width=0.5\linewidth]{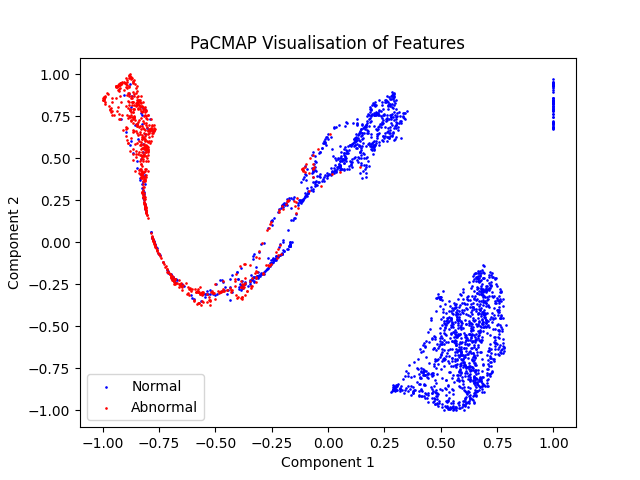}
    \caption{PaCMAP for average 16kHz CinC PCG model.}
    \label{fig:cinc-16k-pacmap}
\end{figure}

\begin{figure}[H]
    \centering
    \includegraphics[width=0.5\linewidth]{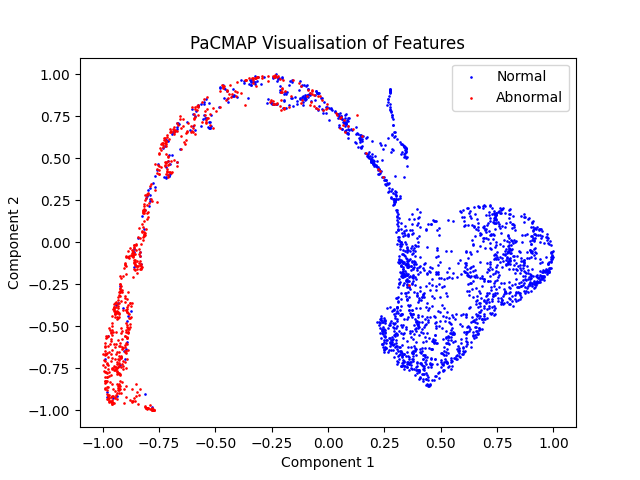}
    \caption{PaCMAP for average 4.125kHz CinC PCG model.}
    \label{fig:cinc-4125-pacmap}
\end{figure}

The PaCMAP~\cite{pacmap} plots of an average-performing model from each experiment are shown below to represent how well, on average, those models encode the data before going through the classifier. The data used for these plots is also from the test set, so these models have not been trained on this data. 
Figure~\ref{fig:cinc-16k-noaugment-pacmap}, Figure~\ref{fig:cinc-16k-pacmap}, and Figure~\ref{fig:cinc-4125-pacmap} illustrate the PaCMAP embeddings for the CinC models, with input data sampled at 16kHz with no augmented dataset, 16kHz with augmented dataset and 4.125kHz with augmented dataset, respectively.


 The augmented models exhibit cleaner class separation than the no-augment baseline, aligning with the higher metrics. The 4.125kHz vs. 16kHz trade-off reflects frequency-band emphasis versus retaining more high-frequency detail.

Table \ref{table:cinc_comparison} shows that this method outperforms current SOTA methods, when trained on all of CinC, with the UAR from the proposed method outperforming all other methods. Although the proposed method does not achieve the highest accuracy, it achieves a greater balance between specificity and sensitivity, resulting in an overall more performant model. This also shows that these methods to create an augmented dataset allow for training of more performant transformer models over the CNNs used in the literature.

\begin{table*}\clearpage
\caption{Models, trained on the entire CinC 2016 dataset, performance comparison with the literature.}
\label{table:cinc_comparison}
\scriptsize
\begin{tabular*}{\textwidth}{@{\extracolsep{\fill}} p{2cm} p{2cm} l l l l l}
\toprule
\textbf{Method} & \textbf{Features} & \textbf{Acc} & \textbf{UAR} & \textbf{TPR} & \textbf{TNR} & \textbf{F1} \\
\midrule
PANNs~\cite{PANNS} & Mel-Spectrogram & -- & 89.70$\pm$1.5\% & 88.60\% & \textbf{96.90\%} & 79.10\% \\
\hline
Dense-FSNet with attention~\cite{dense-fsnet} & Spectrogram-image features & 86.20$\pm$8.42\% & 85.08\% & -- & -- & 84.09\% \\
\hline
Deep CNN~\cite{deepcnn} & Time-domain polynomial chirplet transform & 85.16$\pm$0.49\% & 85.16$\pm$0.49\% & 85.16$\pm$0.49\% & 85.16$\pm$0.49\% & -- \\
\hline
YAMNet~\cite{arnab} & Mel-Spectrogram & \textbf{93.10\%} & 88.31\% & 80.24\% & 96.38\% & 82.53\% \\
\hline
\textbf{This study} & Raw Signal & 92.98$\pm$0.75\% & \textbf{92.48$\pm$1.43\%} & \textbf{93.63$\pm$1.87\%} & 92.48$\pm$0.97\% & \textbf{94.93$\pm$0.50\%} \\
\bottomrule
\end{tabular*}
\end{table*}

\begin{figure}[H]
    \centering
    \begin{subfigure}[t]{0.48\linewidth}
        \centering
        \includegraphics[width=\linewidth]{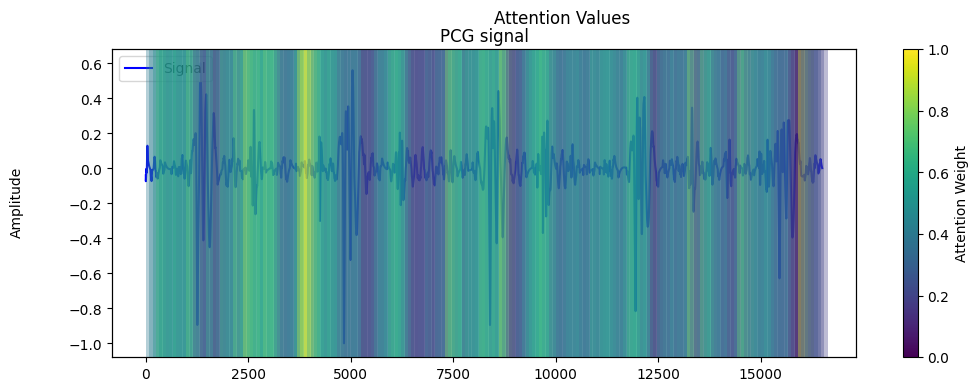}
        \caption{Attention Importance}
        \label{fig:e01791_attention}
    \end{subfigure}
    \hfill
    \begin{subfigure}[t]{0.48\linewidth}
        \centering
        \includegraphics[width=\linewidth]{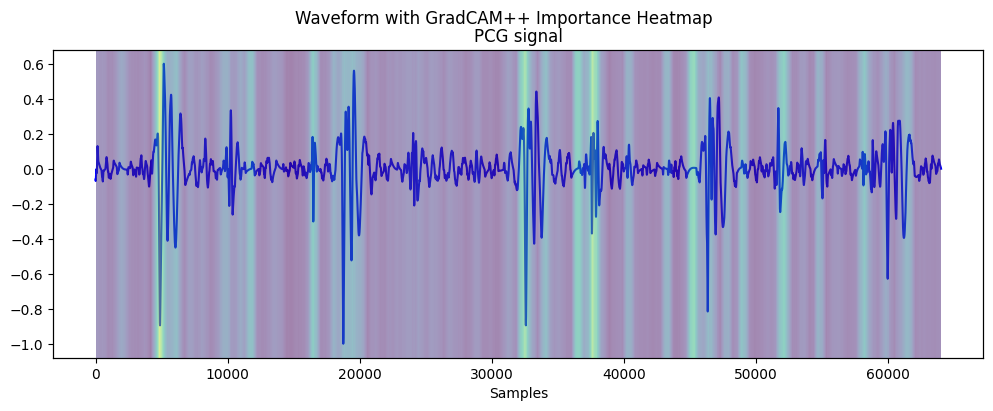}
        \caption{GradCAM++}
        \label{fig:e01791_gradcam}
    \end{subfigure}
    \caption{CinC interpretability images for abnormal subject e01791.}
    \label{fig:e01791_interpretability}
\end{figure}

Figure \ref{fig:e01791_attention} shows the importance from the attention values for each token and how it corresponds to a portion of the signal. Figure \ref{fig:e01791_gradcam} shows the improved gradient-weighted class activation mapping (GradCAM++) importance values from each portion of the signal for the CNN feature encoder. These plots are for an abnormal patient and are representative of plots for all abnormal subjects.

\begin{figure}[H]
    \centering
    \begin{subfigure}[t]{0.48\linewidth}
        \centering
        \includegraphics[width=\linewidth]{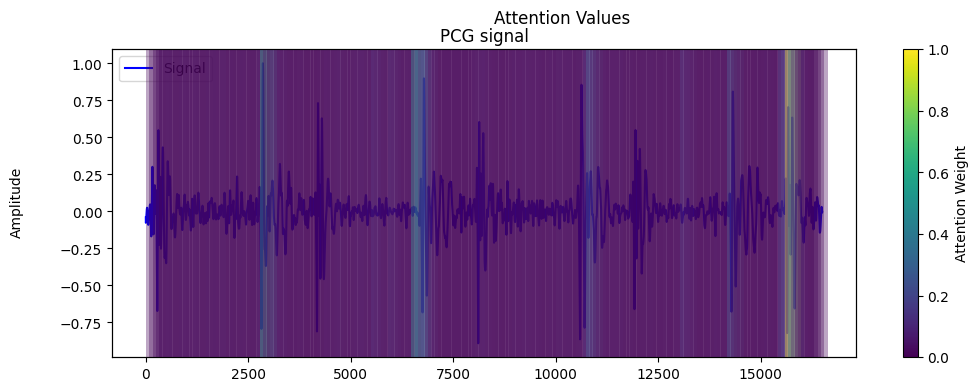}
        \caption{Attention Importance}
        \label{fig:e01427_attention}
    \end{subfigure}
    \hfill
    \begin{subfigure}[t]{0.48\linewidth}
        \centering
        \includegraphics[width=\linewidth]{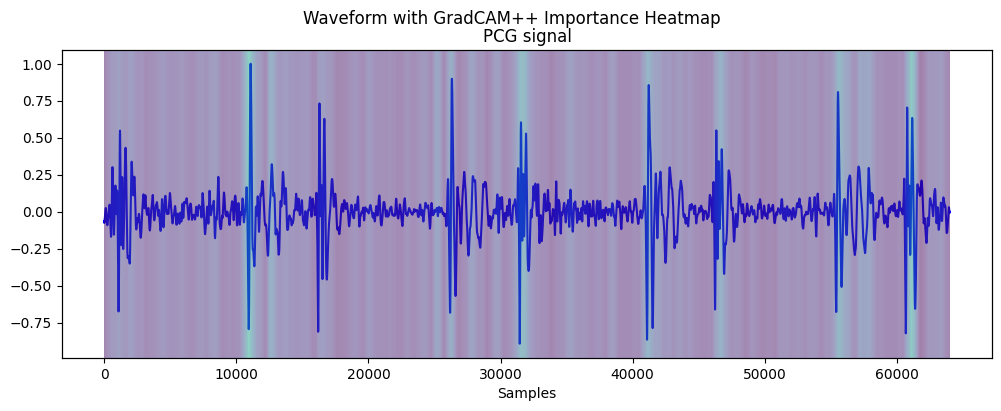}
        \caption{GradCAM++}
        \label{fig:e01427_gradcam}
    \end{subfigure}
    \caption{CinC interpretability images for normal subject e01427.}
    \label{fig:e01427_interpretability}
\end{figure}

Figures \ref{fig:e01427_attention}, \ref{fig:e01427_gradcam}, show the attention importance and GradCAM++ for subject a0352, a normal subject, which is also representative of the feature importance of other normal subjects.

 Interpretability was shown to align with cardiac physiology: abnormal subjects emphasise diastole and systole (murmur-prone phases), while normals emphasise S1/S2—supporting that the model uses clinically relevant features.

\subsection{Multimodal models}

Tables~\ref{table:training_a_results_fragment} and~\ref{table:training_a_results_patient} present the results of models trained on the training-a and augmented datasets.
Each model was trained and tested ten times over five shuffled train-validation-test splits. The results are reported as mean $\pm$ standard deviation.
The best model for each metric is highlighted.

\begin{table}[H]
\captionsetup{skip=10pt}
\caption{Model performance at the fragment level on the training-a dataset.}
\label{table:training_a_results_fragment}
\hspace*{-1cm}
\scriptsize
\begin{tabular}{l l l l l l l l l}
\toprule
\textbf{Inputs} & $f_s$ & \textbf{Data} & \textbf{Acc} & \textbf{UAR} & \textbf{TPR} & \textbf{TNR} & \textbf{F1} & \textbf{MCC} \\
\midrule
PCG+ECG  & 16kHz    & Original    & 71.14$\pm$1.58\% & 50.95$\pm$3.23\% & 99.47$\pm$1.09\% & 2.44$\pm$7.32\% & 83.00$\pm$0.70\% & 0.0252$\pm$0.1046 \\
\hline
PCG+ECG  & 16kHz    & All        & 86.63$\pm$1.89\% & 83.79$\pm$2.63\% & 90.61$\pm$2.31\% & 76.99$\pm$5.54\% & 90.55$\pm$1.35\% & 0.6776$\pm$0.0453 \\
\hline
PCG+ECG  & 4.125kHz & All        & \textbf{90.12$\pm$1.58\%} & 88.08$\pm$2.64\% & \textbf{92.86$\pm$1.62\%} & 83.30$\pm$5.65\% & \textbf{93.07$\pm$1.05\%} & \textbf{0.7603$\pm$0.0430} \\
\bottomrule
\end{tabular}
\end{table}

\begin{table}[H]
\footnotesize
\captionsetup{skip=10pt}
\caption{Model performance at the subject level on the training-a dataset.}
\label{table:training_a_results_patient}
\hspace*{-1cm}
\scriptsize
\begin{tabular}{l l l l l l l l l}
\toprule
\textbf{Inputs} & $f_s$ & \textbf{Data } & \textbf{Acc} & \textbf{UAR} & \textbf{TPR} & \textbf{TNR} & \textbf{F1} & \textbf{MCC} \\
\midrule
PCG+ECG  & 16kHz    & Original  & 70.88$\pm$1.91\% & 50.95$\pm$3.16\% & 99.82$\pm$0.54\% & 2.08$\pm$6.24\% & 82.83$\pm$0.99\% & 0.0322$\pm$0.1229 \\
\hline
PCG+ECG  & 16kHz    & All       & 87.29$\pm$3.28\% & 84.92$\pm$2.86\% & 90.70$\pm$4.70\% & 79.17$\pm$4.56\% & 90.88$\pm$2.54\% & 0.7010$\pm$0.0693 \\
\hline
PCG+ECG  & 4.125kHz & All       & \textbf{93.14$\pm$1.80\%} & \textbf{92.21$\pm$2.58\%} & \textbf{94.35$\pm$2.60\%} & 90.10$\pm$5.88\% & \textbf{95.12$\pm$1.29\%} & \textbf{0.8380$\pm$0.0436} \\
\bottomrule
\end{tabular}
\end{table}

 Without augmentation, fine-tuning collapses (UAR $\approx$ 51\%, TNR $\approx$ 2\%). Augmentation is shown to enable effective training of the transformer-based Wav2Vec2 Encoder model. Resampling the inputs to 4.125kHz yields the strongest overall metrics, indicating a practical benefit to bandwidth-matched inputs. 

\begin{figure}[H]
  \centering
  \begin{subfigure}[b]{0.45\textwidth}
    \includegraphics[width=\textwidth]{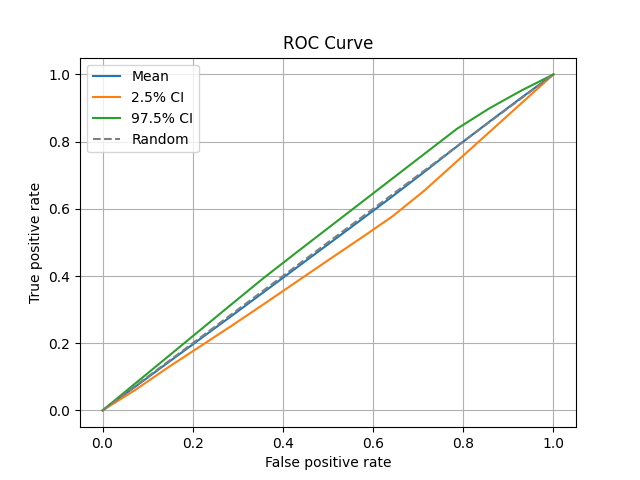}
    \caption{Fragment level.}
  \end{subfigure}
  \hfill
  \begin{subfigure}[b]{0.45\textwidth}
    \includegraphics[width=\textwidth]{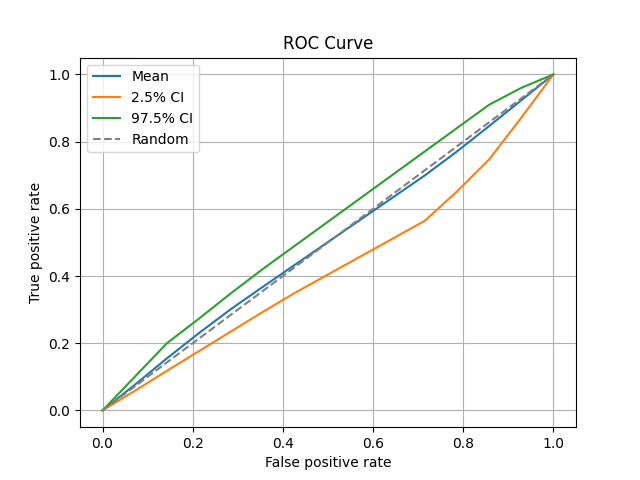}
    \caption{Subject level.}
  \end{subfigure}
  \caption{PCG+ECG 16kHz no augments training-a model ROC plots}
  \label{fig:ta-pecg-16k-noaug}
\end{figure}

\begin{figure}[H]
  \centering
  \begin{subfigure}[b]{0.45\textwidth}
    \includegraphics[width=\textwidth]{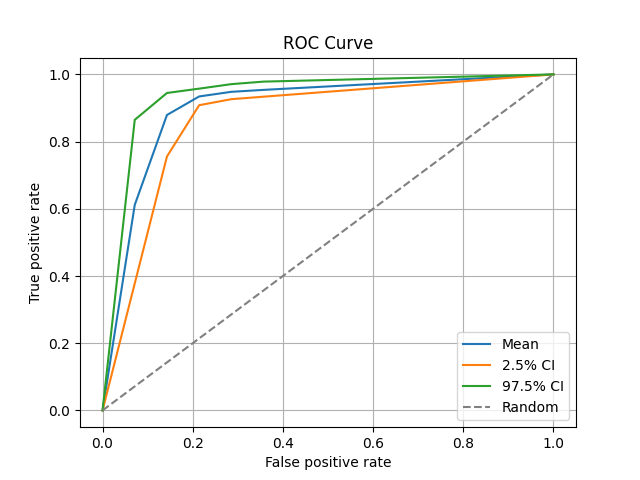}
    \caption{Fragment level.}
  \end{subfigure}
  \hfill
  \begin{subfigure}[b]{0.45\textwidth}
    \includegraphics[width=\textwidth]{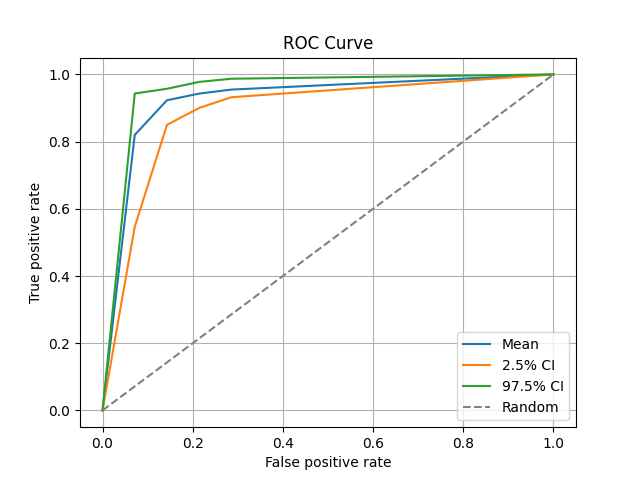}
    \caption{Subject level.}
  \end{subfigure}
  \caption{PCG+ECG 16kHz training-a model ROC plots}
  \label{fig:ta-pecg-16k}
\end{figure}

\begin{figure}[H]
  \centering
  \begin{subfigure}[b]{0.45\textwidth}
    \includegraphics[width=\textwidth]{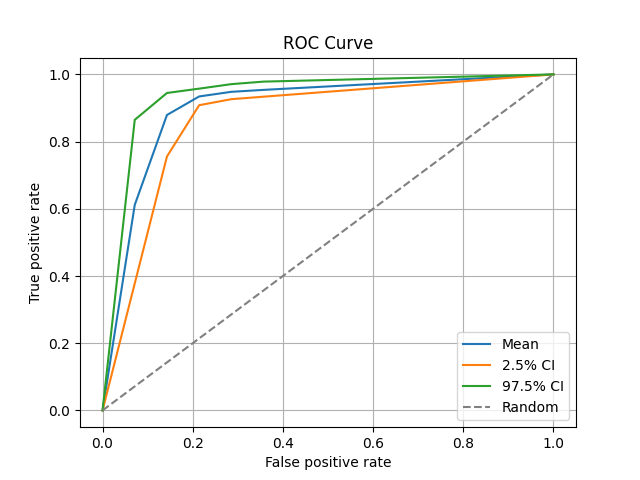}
    \caption{Fragment level.}
  \end{subfigure}
  \hfill
  \begin{subfigure}[b]{0.45\textwidth}
    \includegraphics[width=\textwidth]{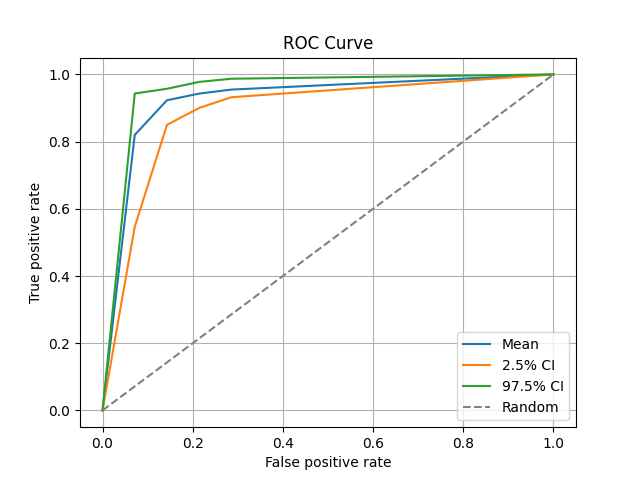}
    \caption{Subject level.}
  \end{subfigure}
  \caption{PCG+ECG 4.125kHz training-a model ROC plots}
  \label{fig:ta-pecg-4125}
\end{figure}

The ROC curves of the multimodal models are in Figure~\ref{fig:ta-pecg-16k-noaug},  Figure~\ref{fig:ta-pecg-16k} and Figure~\ref{fig:ta-pecg-4125} for the multimodal models with no augments at 16kHz, with augments and 16kHz, and with augments and 4.125kHz, respectively.

 ROC curves confirm that augmentation substantially improves the operating range; the non-augmented multimodal model exhibits pronounced threshold sensitivity.

\begin{figure}[H]
    \centering
    \includegraphics[width=0.5\linewidth]{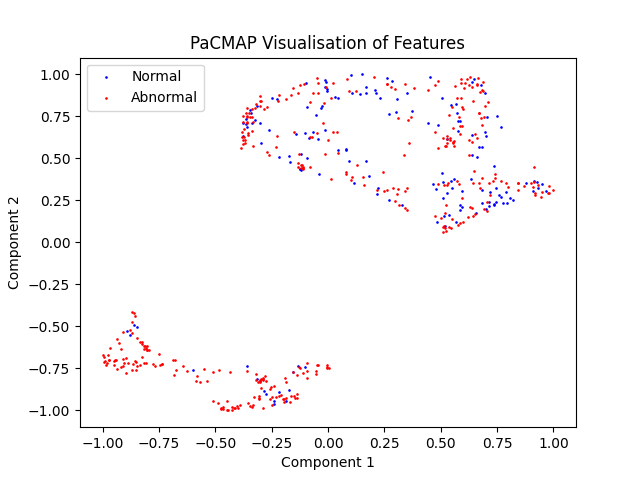}
    \caption{PaCMAP for average no augment 16kHz multimodal model.}
    \label{fig:pecg-16k-noaug-pacmap}
\end{figure}

\begin{figure}[H]
    \centering
    \includegraphics[width=0.5\linewidth]{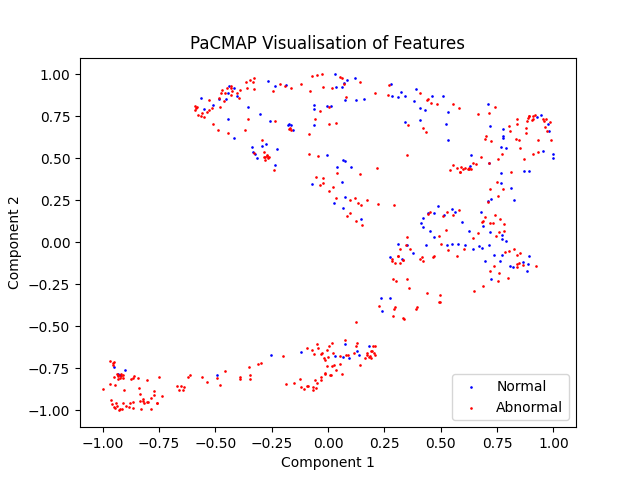}
    \caption{PaCMAP for average 16kHz multimodal model.}
    \label{fig:pecg-16k-pacmap}
\end{figure}

\begin{figure}[H]
    \centering
    \includegraphics[width=0.5\linewidth]{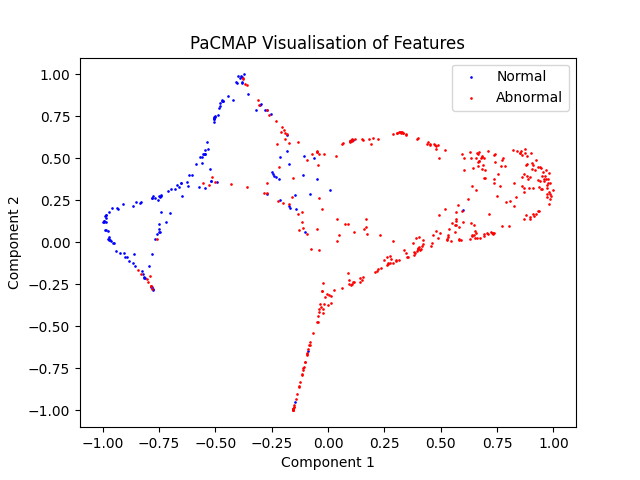}
    \caption{PaCMAP for average 4.125kHz multimodal model.}
    \label{fig:pecg-4125-pacmap}
\end{figure}

The multimodal PaCMAP plots are in Figure~\ref{fig:pecg-16k-noaug-pacmap}, Figure~\ref{fig:pecg-16k-pacmap} and Figure~\ref{fig:pecg-4125-pacmap}, for the multimodal models with no augments at 16kHz, with augments at 16kHz, and with augments at 4.125kHz, respectively.


 Augmentation tightens clusters markedly; 4.125kHz shows the clearest separation, consistent with the best quantitative metrics.

Comparing our model that utilises both PCG and ECG signals to other models that were evaluated on training-a,
We can see that our model has the greatest performance with the highest for each metric, other than TPR, in
Table \ref{table:training_a_comparison}.
Although it does not achieve the highest TPR, our proposed model achieves a much higher TNR, than the CNN model from \cite{milan}, resulting in all other metrics being greater and hence a better performing model.
It is observed that the majority of methods within this comparison utilise the raw signal, as opposed to spectrograms, which is more common for models trained on the entire CinC dataset.
However, none of the previous approaches utilised transformer architectures.
As training-a is a small dataset, and hence, models trained on it will not generalise well, the results from training on the CinC dataset demonstrate that the method can effectively scale to larger amounts of data while still delivering performance improvements.

\begin{table*}\clearpage
\caption{Models, trained on CinC 2016 training-a dataset, performance comparison with the literature.}
\label{table:training_a_comparison}
\scriptsize
\begin{tabular*}{\textwidth}{@{\extracolsep{\fill}} l l l l l l l l}
\toprule
\textbf{Method} & \textbf{Features} & \textbf{Acc} & \textbf{UAR} & \textbf{TPR} & \textbf{TNR} & \textbf{F1} \\
\midrule
CNN-SVM~\cite{cnnsvm} & Raw Signal  & 87.30$\pm$1.00\% & 87.40$\pm$1.20\% & 90.30$\pm$0.60\% & 84.50$\pm$1.80\% & 87.40$\pm$1.00\%  \\
\hline
CNN~\cite{milan}      & Spectrogram & 91.25\%          & 84.17\%          & \textbf{98.33\%} & 70.00\%          & 94.40\%           \\
\hline
RNN~\cite{mattone}    & Raw Signal  & 91.60\%          & 91.55\%          & 92.00\%          & 91.10\%          & 91.50\%           \\
\hline
\textbf{This study}   & Raw Signal  & \textbf{93.14$\pm$1.80\%} & \textbf{92.21$\pm$2.58\%} & 94.35$\pm$2.60\% & \textbf{90.10$\pm$5.88\%} & \textbf{95.12$\pm$1.29\%} \\
\bottomrule
\end{tabular*}
\end{table*}

 Prior SOTA relies on CNN/RNN backbones; the highest TPR (CNN) comes with low TNR (70\%), i.e., false positives. The proposed multimodal transformer with augmentation delivers the best overall performance, balancing TPR/TNR.

\begin{figure}[H]
    \centering
    \begin{subfigure}[t]{0.48\linewidth}
        \centering
        \includegraphics[width=\linewidth]{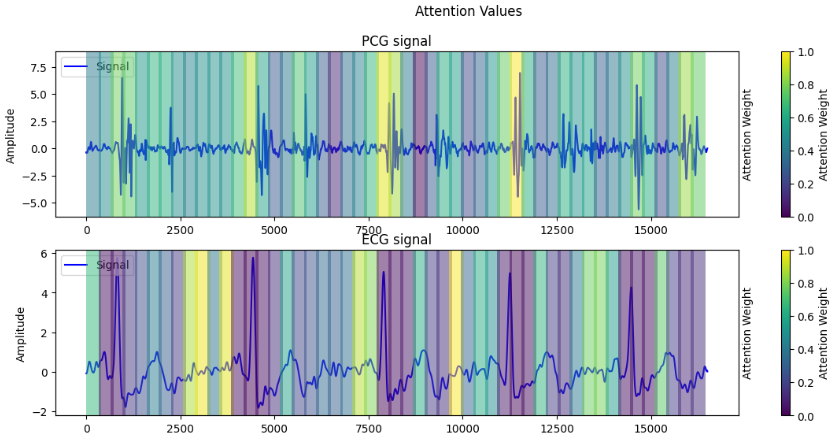}
        \caption{Attention Importance}
        \label{fig:a0327_attention}
    \end{subfigure}
    \hfill
    \begin{subfigure}[t]{0.48\linewidth}
        \centering
        \includegraphics[width=\linewidth]{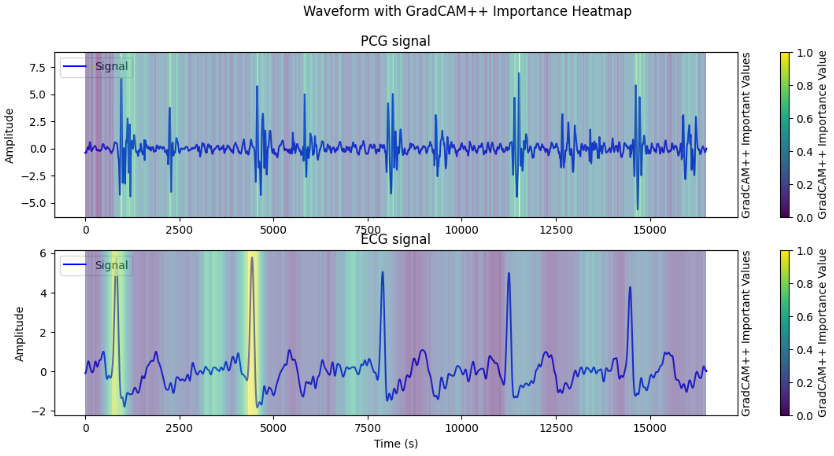}
        \caption{GradCAM++}
        \label{fig:a0327_gradcam}
    \end{subfigure}
    \caption{CinC interpretability images for abnormal subject a0327.}
    \label{fig:a0327_interpretability}
\end{figure}

Figure \ref{fig:a0327_attention} shows the importance from the attention values for each token and how it corresponds to a portion of the signal, with Figure \ref{fig:a0327_gradcam} showing the GradCAM++ importance values from each portion of the signal for the CNN feature encoder. These plots are for an abnormal patient and are representative of plots for all abnormal subjects.

\begin{figure}[H]
    \centering
    \begin{subfigure}[t]{0.48\linewidth}
        \centering
        \includegraphics[width=\linewidth]{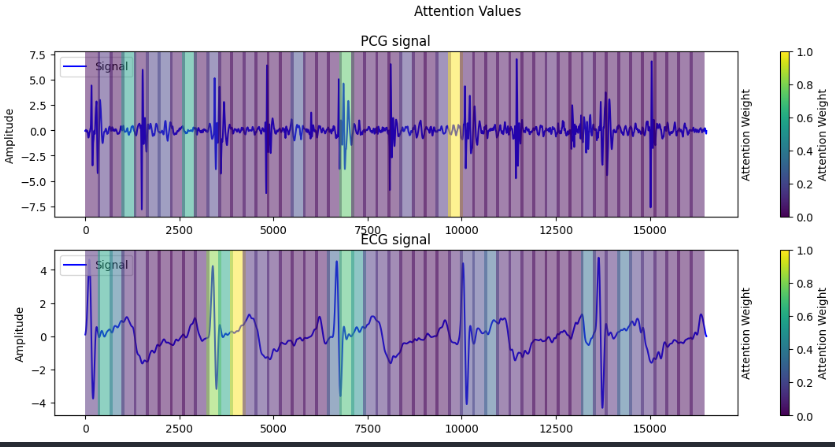}
        \caption{Attention Importance}
        \label{fig:a0352_attention}
    \end{subfigure}
    \hfill
    \begin{subfigure}[t]{0.48\linewidth}
        \centering
        \includegraphics[width=\linewidth]{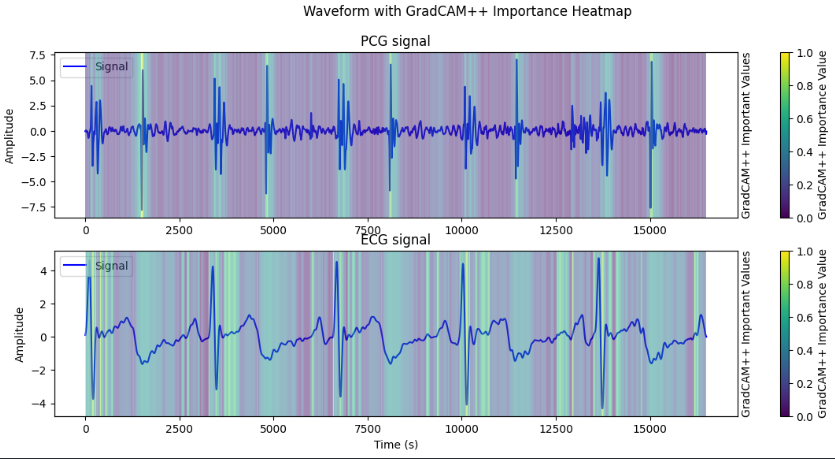}
        \caption{GradCAM++}
        \label{fig:a0352_gradcam}
    \end{subfigure}
    \caption{CinC interpretability images for normal subject a0352.}
    \label{fig:a0352_interpretability}
\end{figure}

Figures \ref{fig:a0352_attention}, \ref{fig:a0352_gradcam}, show the attention importance and GradCAM++ for subject a0352, a normal subject, which is also representative of the feature importance of other normal subjects.

Similar to the single model case, attention focuses on diastolic/systolic phases for abnormal subjects and S1/S2 for normals, indicating clinically sensible feature usage. The features from the ECG signal were also found to align with morphologically consistent features.

\subsection{Multichannel models}

For the vest dataset, a seven-fold cross-validation was utilised. Each experiment was run ten times over three shuffles of the seven-folds. Table~\ref{table:vest_data_fragment} and Table~\ref{table:vest_data_patient} show the fragment-level and subject-level results of the experiments on the vest dataset.

\begin{table}[H]
\footnotesize
\captionsetup{skip=10pt}
\caption{Model performance at the fragment level on the vest dataset.}
\label{table:vest_data_fragment}
\scriptsize
\begin{tabular}{l l l l l l l l l}
\toprule
$f_s$ & \textbf{Data} & \textbf{Acc} & \textbf{UAR} & \textbf{TPR} & \textbf{TNR} & \textbf{F1} & \textbf{MCC} \\
\midrule
16kHz & Original        & $64.41\pm0.05\%$ & $59.05\pm0.05\%$ & \bm{$82.33\pm0.12\%$}  & $35.77\pm0.03\%$  & $73.95\pm0.06\%$  &  $0.2049\pm0.0011$ \\
\hline
16kHz & All             & $68.74\pm0.22\%$ & $66.61\pm0.25\%$ & $76.27\pm0.18\%$ & $56.96\pm0.38\%$ & $74.93\pm0.17\%$ & $0.3368\pm0.0048$ \\
\hline
4.125kHz & All          & \bm{$70.66\pm0.21\%$}  & \bm{$68.23\pm0.18\%$} & $78.88\pm0.46\%$ & \bm{$57.57\pm0.30\%$} & \bm{$76.78\pm0.22\%$} & \bm{$0.3715\pm0.0044$} \\
\hline
\bottomrule
\end{tabular}
\end{table}

\begin{table}[H]
\footnotesize
\captionsetup{skip=10pt}
\caption{Model performance at the subject level on the vest dataset.}
\label{table:vest_data_patient}
\scriptsize
\begin{tabular}{l l l l l l l l l}
\toprule
$f_s$ & \textbf{Data} & \textbf{Acc} & \textbf{UAR} & \textbf{TPR} & \textbf{TNR} & \textbf{F1} & \textbf{MCC} \\
\midrule
16kHz & Original        & $67.33\pm0.51\%$ & $61.67\pm0.50\%$ & $85.38\pm1.00\%$  & $37.96\pm0.80\%$ & $73.95\pm0.06\%$ & $0.2824\pm0.0136$ \\
\hline
16kHz & All             & $73.84\pm0.56\%$ & $71.99\pm0.59\%$ & $80.52\pm0.70\%$ & $63.46\pm1.01\%$ & $79.08\pm0.48\%$ & $0.4495\pm0.0124$ \\
\hline
4.125kHz & All          & \bm{$77.13\pm1.50\%$}  & \bm{$74.25\pm1.73\%$} & \bm{$86.47\pm1.30\%$} & \bm{$62.04\pm2.76\%$} & \bm{$82.34\pm1.10\%$} & \bm{$0.5082\pm0.0345$} \\
\bottomrule
\end{tabular}
\end{table}

 Including augmentation substantially boosts performance; 9.8\% increase in accuracy and 12.58\% increase to UAR (subject-level), with TNR increases $\sim$+24.08\%, reducing positive bias on this challenging set. LoRA and the addition of the SVM help train effectively with limited data, yet the augmented dataset still provides significant additional gains.

\begin{figure}[H]
  \centering
  \begin{subfigure}[b]{0.45\textwidth}
    \includegraphics[width=\textwidth]{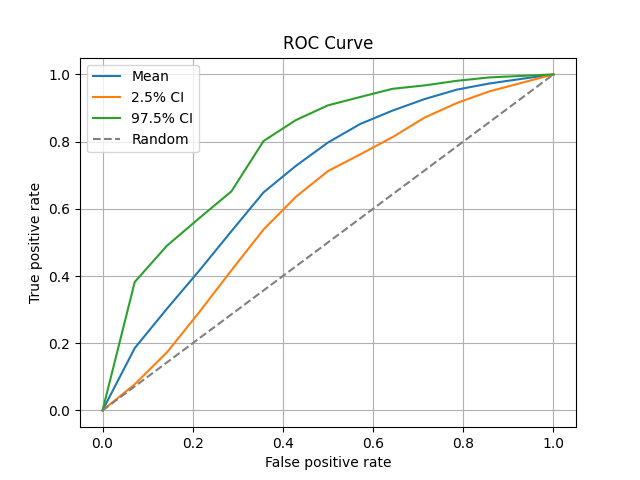}
    \caption{Fragment level.}
  \end{subfigure}
  \hfill
  \begin{subfigure}[b]{0.45\textwidth}
    \includegraphics[width=\textwidth]{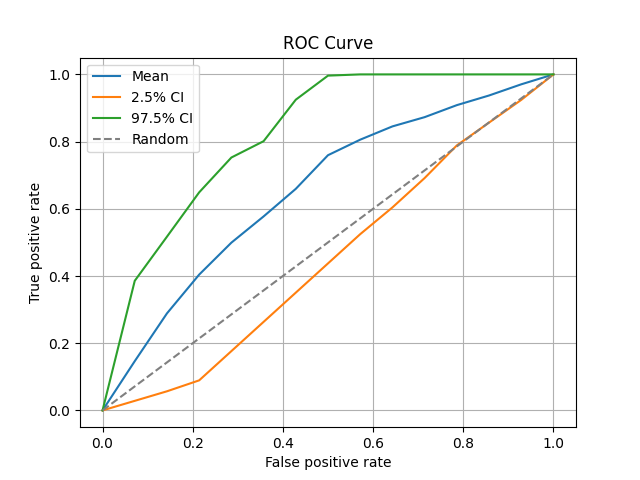}
    \caption{Subject level.}
  \end{subfigure}
  \caption{mPCG 16kHz no augments vest dataset model ROC plots}
  \label{fig:vest-noaug-16k}
\end{figure}

\begin{figure}[H]
  \centering
  \begin{subfigure}[b]{0.45\textwidth}
    \includegraphics[width=\textwidth]{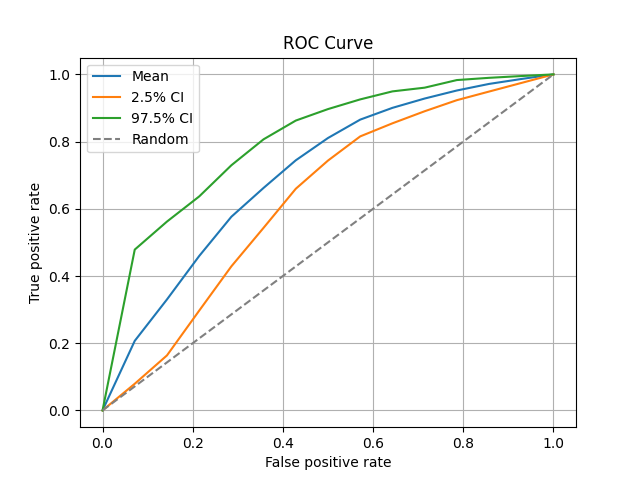}
    \caption{Fragment level.}
  \end{subfigure}
  \hfill
  \begin{subfigure}[b]{0.45\textwidth}
    \includegraphics[width=\textwidth]{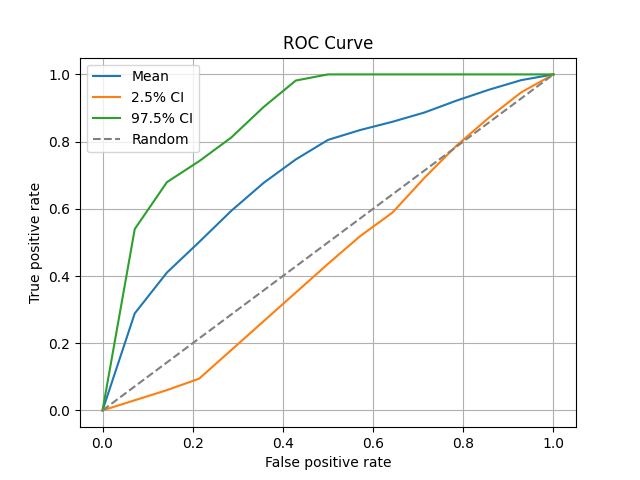}
    \caption{Subject level.}
  \end{subfigure}
  \caption{mPCG 16kHz vest dataset model ROC plots}
  \label{fig:vest-16k}
\end{figure}

\begin{figure}[H]
  \centering
  \begin{subfigure}[b]{0.45\textwidth}
    \includegraphics[width=\textwidth]{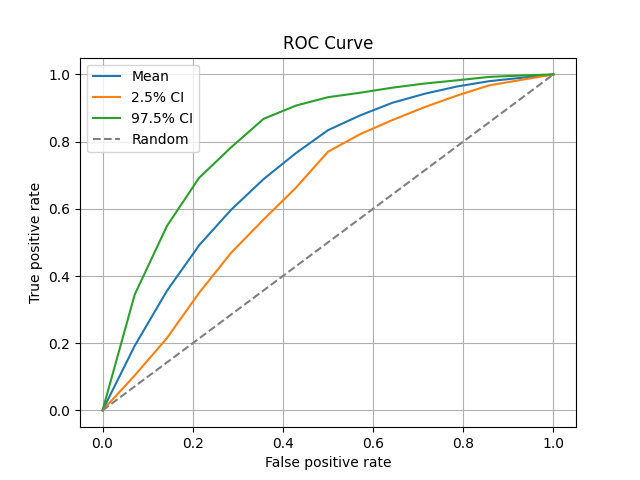}
    \caption{Fragment level.}
  \end{subfigure}
  \hfill
  \begin{subfigure}[b]{0.45\textwidth}
    \includegraphics[width=\textwidth]{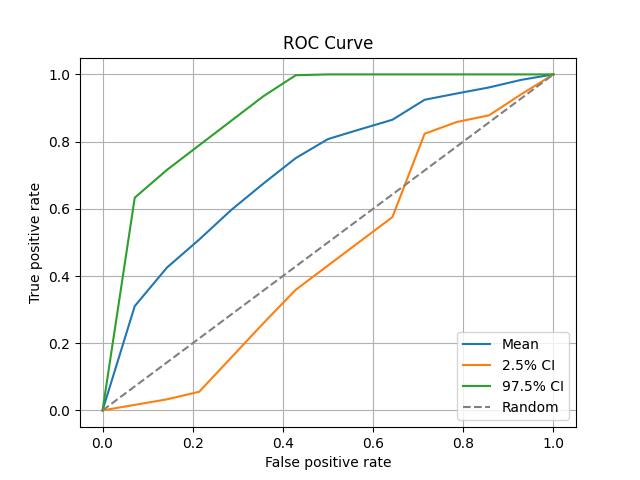}
    \caption{Subject level.}
  \end{subfigure}
  \caption{mPCG 4.125kHz vest dataset model ROC plots}
  \label{fig:vest-4125}
\end{figure}

Figure~\ref{fig:vest-noaug-16k}, Figure~\ref{fig:vest-16k} and Figure~\ref{fig:vest-4125}, convey the ROC curve for the multichannel models with inputs sampled at 16kHz with no augmented dataset, 16kHz and 4.125kHz respectively.

 ROC areas are lower than CinC and training-a (as expected for low-intensity murmurs), but the augmented models present more favourable operating regions.

\begin{figure}[H]
    \centering
    \includegraphics[width=0.5\linewidth]{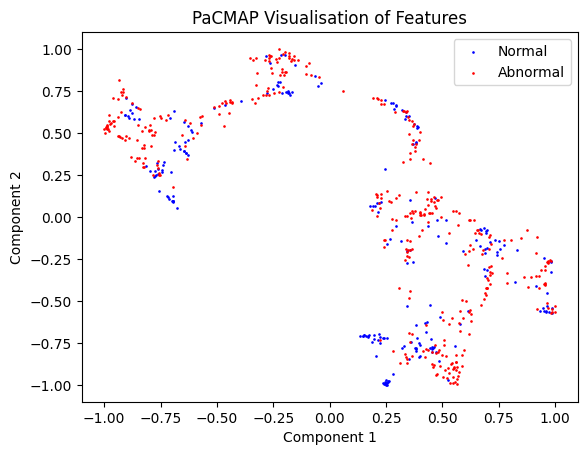}
    \caption{PaCMAP for average 16kHz mPCG model.}
    \label{fig:vest-16k-pacmap}
\end{figure}

\begin{figure}[H]
    \centering
    \includegraphics[width=0.5\linewidth]{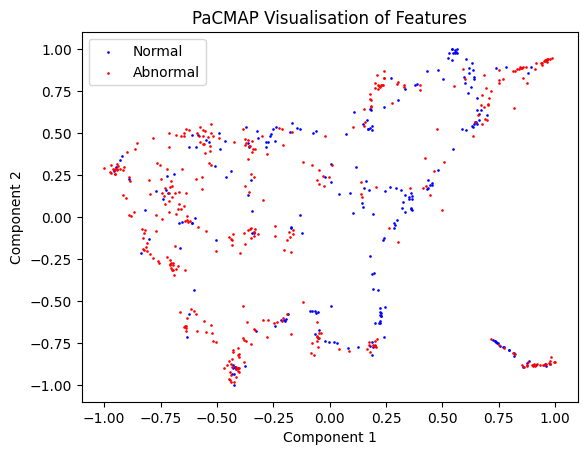}
    \caption{PaCMAP for average no augment 16kHz mPCG model.}
    \label{fig:vest-16k-noaugment-pacmap}
\end{figure}

\begin{figure}[H]
    \centering
    \includegraphics[width=0.5\linewidth]{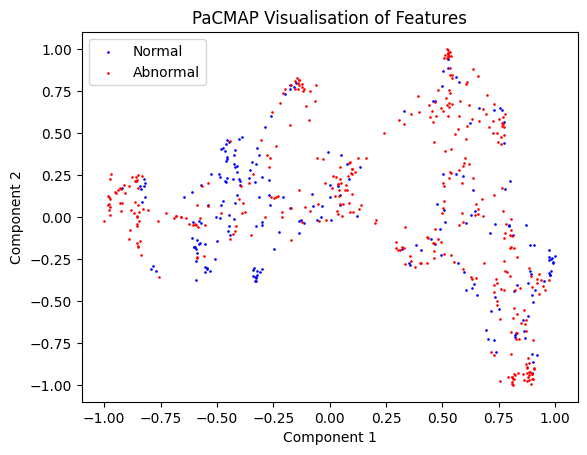}
    \caption{PaCMAP for average 4.125kHz mPCG model.}
    \label{fig:vest-4125-pacmap}
\end{figure}

The vest datasets models' embeddings are shown in Figure~\ref{fig:vest-16k-pacmap}, Figure~\ref{fig:vest-16k-noaugment-pacmap} and Figure~\ref{fig:vest-4125-pacmap}, representing the models with inputs sampled at 16kHz, 16kHz and no augmented dataset and 4.125kHz, respectively.


 Embedding separation is weaker than in CinC and training-a, consistent with dataset difficulty; nonetheless, augmentation improves clustering, which aligns with the higher metrics.

\begin{figure}[H]
    \centering
    \begin{subfigure}[t]{0.48\linewidth}
        \centering
        \includegraphics[width=\linewidth]{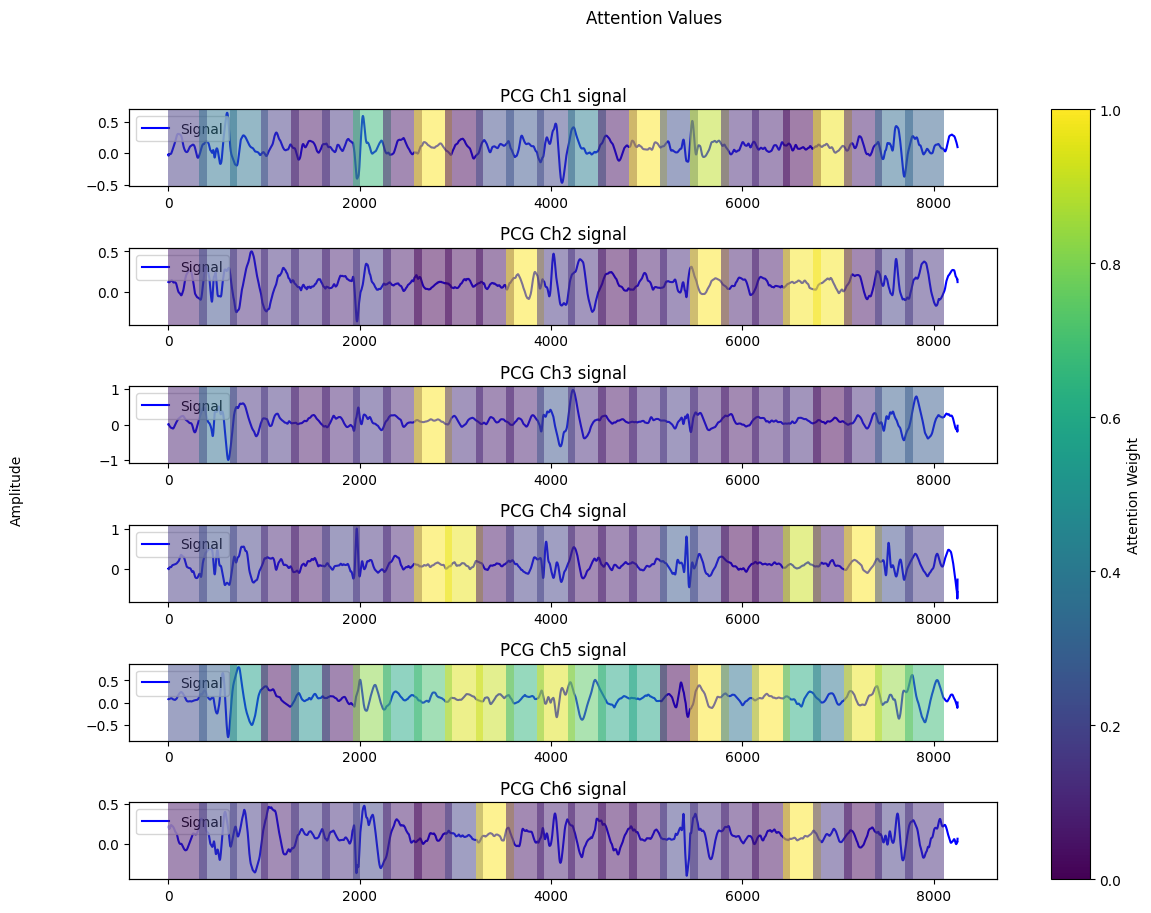}
        \caption{Attention Importance}
        \label{fig:cad_attention}
    \end{subfigure}
    \hfill
    \begin{subfigure}[t]{0.48\linewidth}
        \centering
        \includegraphics[width=\linewidth]{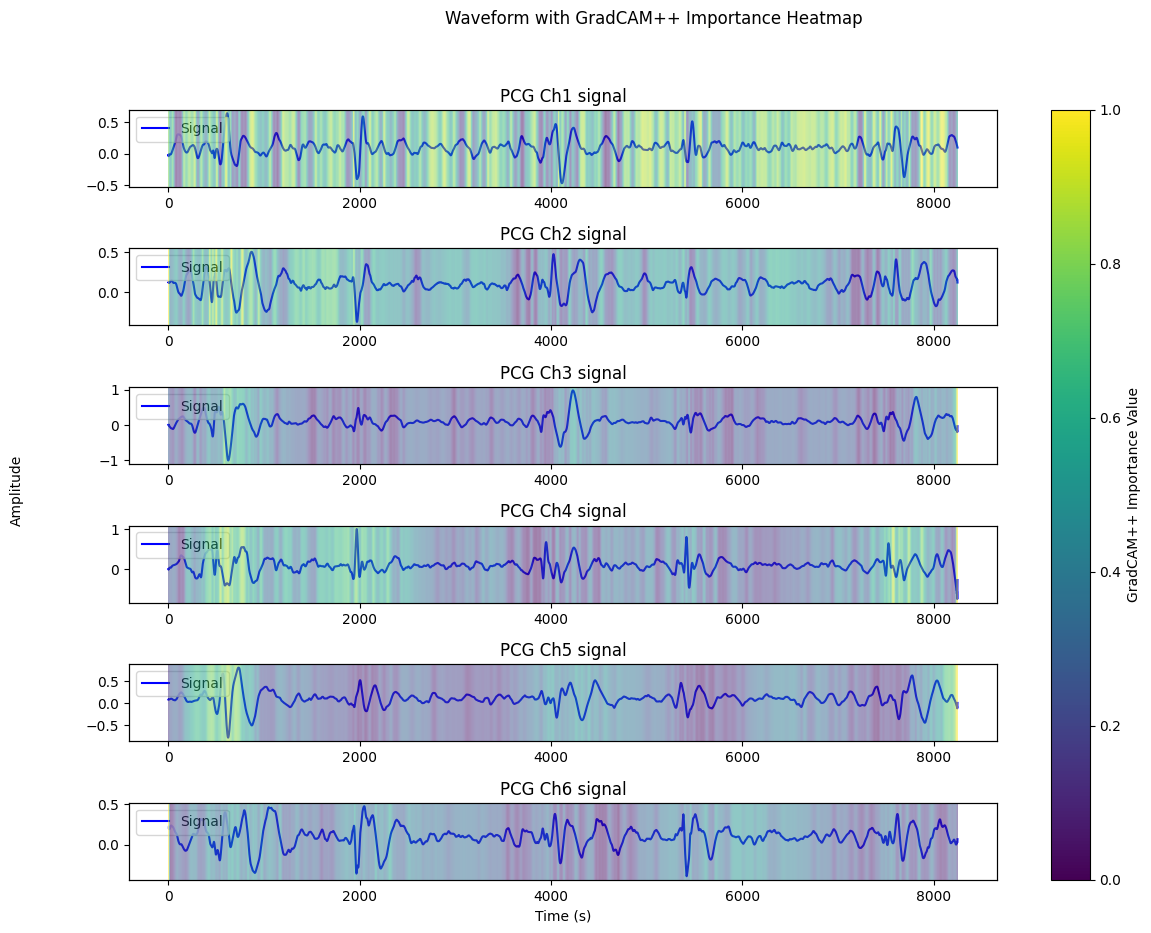}
        \caption{GradCAM++}
        \label{fig:cad_gradcam}
    \end{subfigure}
    \caption{Vest data interpretability images for abnormal CAD subject.}
    \label{fig:cad_interp}
\end{figure}

\begin{figure}[H]
    \centering
    \begin{subfigure}[t]{0.48\linewidth}
        \centering
        \includegraphics[width=\linewidth]{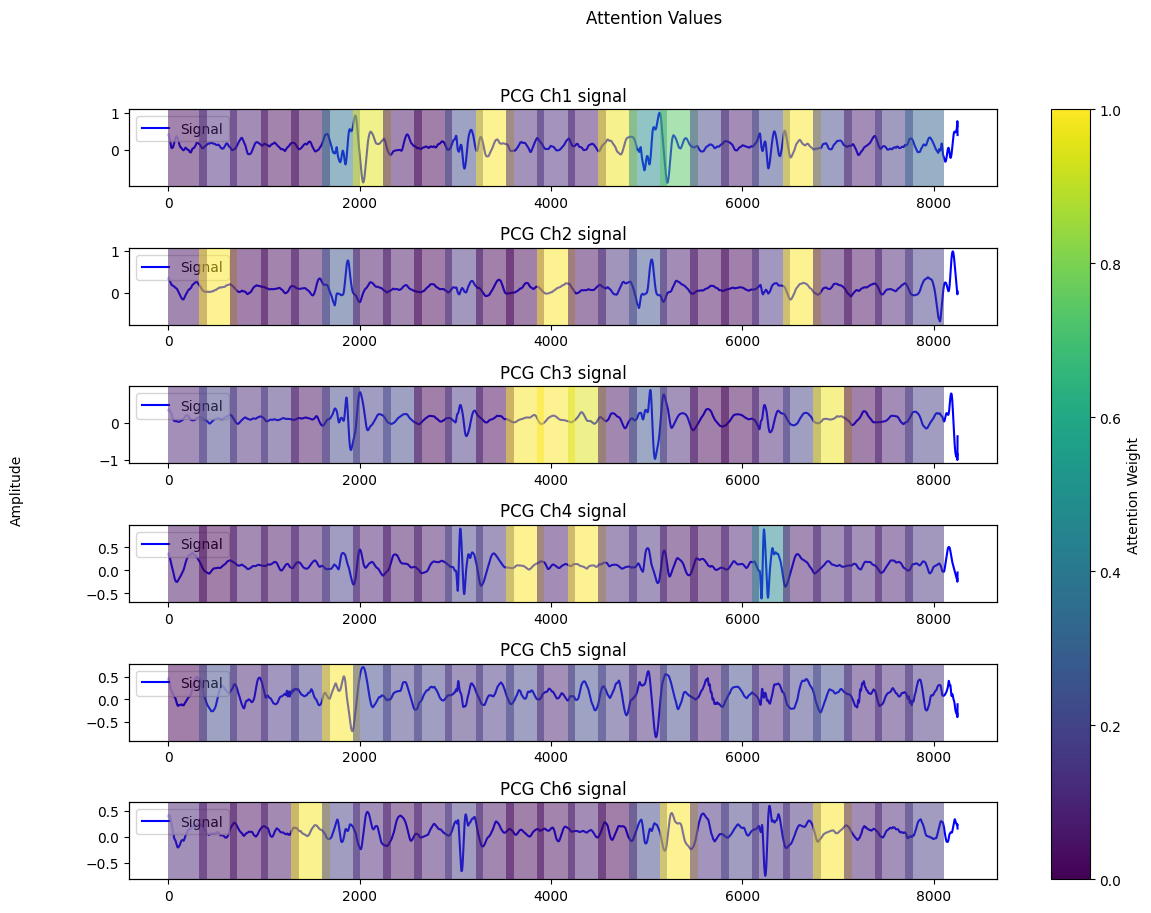}
        \caption{Attention Importance}
        \label{fig:nor_attention}
    \end{subfigure}
    \hfill
    \begin{subfigure}[t]{0.48\linewidth}
        \centering
        \includegraphics[width=\linewidth]{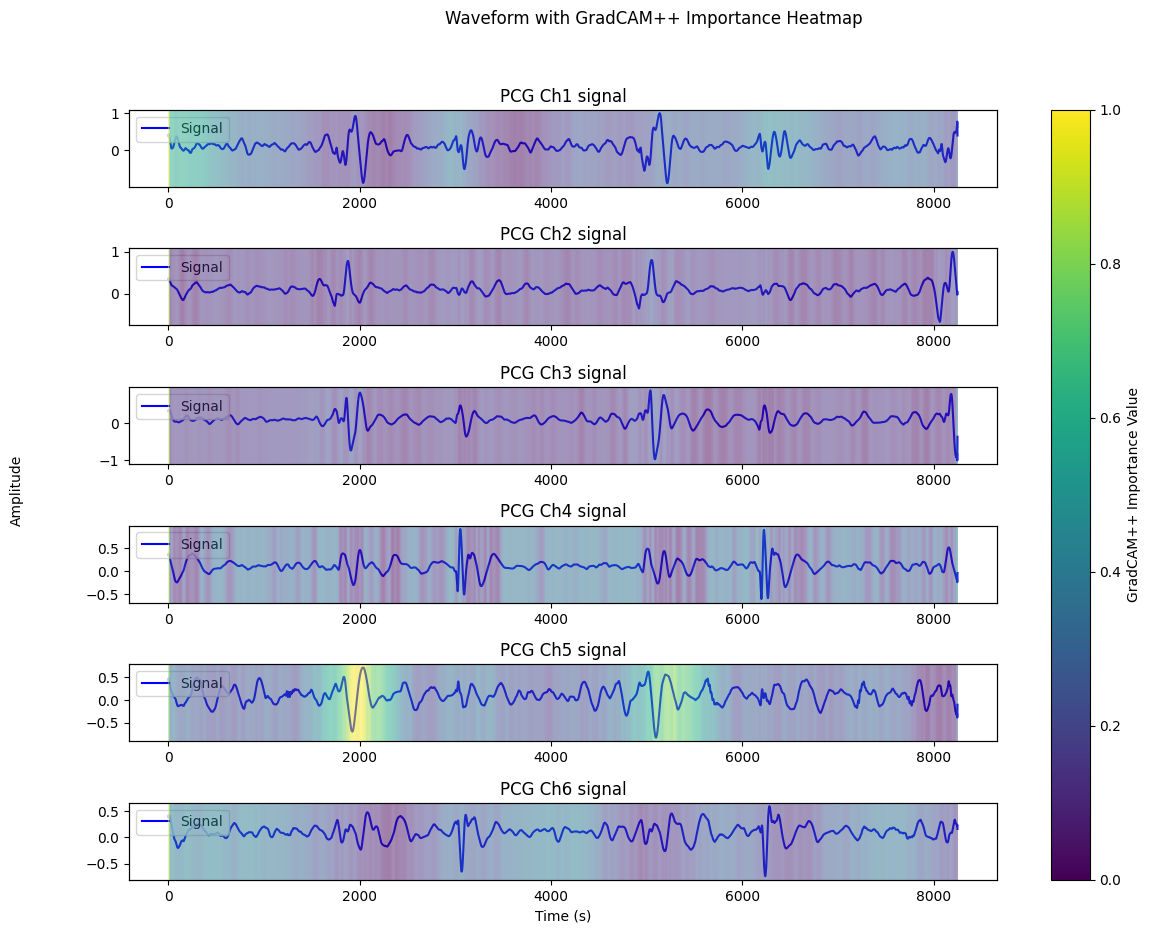}
        \caption{GradCAM++}
        \label{fig:nor_gradcam}
    \end{subfigure}
    \caption{Vest data interpretability images for a normal subject.}
    \label{fig:nor_interpretability}
\end{figure}

The attention importance and GradCAM++ for a CAD subject in the vest dataset are illustrated in Figure~\ref{fig:cad_interp}. Figure~\ref{fig:nor_interpretability} demonstrates the attention importance and GradCAM++ for a normal subject of the same dataset.

 Both attention and GradCAM++ emphasise diastolic regions characteristic of CAD murmurs; this remains true in noisy conditions, indicating the model attends to clinically meaningful cues. Further, it is seen that certain channels have larger contributions than others for certain subjects, showing that it is useful to combine channels.

Comparing with the method from the original study~\cite{vestpaper}, it was observed that their LFCC--SVM approach slightly outperforms the proposed method when evaluated only on the small, breath-held dataset collected from a single round of data acquisition. However, when the original method is re-run on the larger dataset used in this study—which includes an additional round of recordings collected in a different hospital ward—the proposed method achieves superior overall performance. The LFCC--SVM baseline also benefits from cleaner, hand-segmented data recorded while subjects held their breath, whereas the proposed method is trained and evaluated on free-breathing recordings that contain respiratory and background noise, resulting in a noticeable distribution shift relative to the first collection round. As shown in Table~\ref{tab:vestcomp}, the proposed method provides strong improvements in several clinically relevant metrics, most notably achieving the highest TPR (86.47\%) and F1-score (82.34\%), indicating improved sensitivity to CAD cases and better balance between precision and recall. Although the LFCC--SVM baseline maintains a higher TNR, its performance degrades substantially when extended to the larger dataset, while the proposed method remains stable under noisier real-world conditions. These findings highlight the improved robustness and practicality of the proposed method, which does not require manual heart-cycle segmentation or controlled breath-holding and is therefore better suited for deployment in realistic clinical environments.

\begin{table*}\clearpage
\caption{Models, trained on the vest data dataset, performance comparison with the literature\\ \footnotesize{Note: models use different datasets, both collected from the same hospital with the same hardware.}}
\label{tab:vestcomp}
\scriptsize
\begin{tabular*}{\textwidth}{@{\extracolsep{\fill}} p{1cm}p{2.5cm}p{0.9cm}llllll }
\toprule
\textbf{Method} & \textbf{Data} &  \textbf{Features} & \textbf{Acc} & \textbf{UAR} & \textbf{TPR} & \textbf{TNR} & \textbf{F1} \\
\midrule
LFCC SVM~\cite{vestpaper} & 80 subjects (40CAD/40NOR), 10s breath held & LFCCs  & \textbf{80.44\%} & \textbf{80.435\%} & 85.25\% & \textbf{75.62\%} & 81.00\% \\
\hline
\hline
LFCC SVM~\cite{vestpaper} & 157 subjects (96CAD/61NOR) 10s breath held & LFCCs & 
73.37$\pm$0.79\% & 73.21$\pm$1.01\% & 76.77$\pm$1.08\% & 69.64$\pm$0.92\% & 75.08$\pm$0.71\% \\
\hline
\textbf{This study}   & 157 subjects (96CAD/61NOR) 60s free breathing &  Raw Signal  & 77.13$\pm$1.50\% & 74.25$\pm$1.73\% & \textbf{86.47$\pm$1.30\%} & 62.04$\pm$2.76\% & \textbf{82.34$\pm$1.10\%}  \\
\bottomrule
\end{tabular*}
\end{table*}

\subsection{Model and dataset performance comparison}

The different datasets saw different performances as expected due to the differing difficulty in classification, the quality and quantity of the data, and the diseases represented in the dataset. 
Slight modifications were made to the procedure with very limited datasets, as in the case of the vest dataset, where an SVM was used along with LoRA to improve performance further. The single channel dataset represented a reasonably sized dataset, the multimodal dataset represented a smaller dataset, and the multichannel vest dataset represented a small real-world dataset. However, in all cases, the use of the augmented and synthetic data further improved the performance of the classifiers across all scenarios. This demonstrates the usefulness of the proposed method in the classification of abnormal heart sounds using single channel,  multimodal and multichannel data.

\section{Conclusion and future work}
\label{sec:conclusion}
This paper demonstrates the usefulness of the scalable architecture and augmentation procedure for single-channel PCG data, multimodal PCG and ECG and mPCG data for the classification of abnormal heart sounds. The approach achieves SOTA performance. 
On the CinC 2016 dataset of single-channel PCG, accuracy, UAR, sensitivity, specificity, and MCC reach 92.48\%, 93.05\%, 93.63\%, 92.48\%, and 0.8283, respectively. 
Using the synchronised PCG and ECG signals of the CinC 2016 training-a dataset, the method achieves 93.14\% accuracy, 92.21\% UAR, 94.35\% sensitivity, 90.10\% specificity, and 0.8380 MCC. 
On a wearable vest dataset consisting of mPCG signals, the model attains 77.13\% accuracy, 74.25\% UAR, 86.47\% sensitivity, 62.04\% specificity, and 0.5082 MCC. 
The method is also validated on real-world data from this multichannel dataset, though the amount of data collected remains limited.  

Further work is needed to improve the multichannel diffusion model to train and generate on all channels at once, as well as further investigation into the potentially disease-specific augmentations, and to try to find the optimal ordering and use of augmentations. Additionally, further work will include testing the methodology on larger real-world datasets to assess how well it scales to larger datasets.

\section{Ethics approval and consent}
This study received approval from the ethics committee of Fortis Hospital, Kolkata, India, where the data collection took place (ECR/240/Inst/WB/2013/RR-19, Date of approval: 13/01/2023). Informed consent was obtained from all subjects, and data collection adhered to the code of ethics for conducting research on human subjects as outlined in the Helsinki Declaration.

\section{Funding}

This research did not receive any specific grant from funding agen-
cies in the public, commercial, or not-for-profit sectors.

\section{Acknowledgement}
We thank Ticking Heart Pty. Ltd. for providing their wearable vest data. We thank Harry Walters for his invaluable feedback and comments over coffee in the kitchen. We also thank Danny Baker, Enzo and Tank for their encouragement and support.

\section{Code Availability}
All code can be found in the following location \url{https://github.com/MilanMarocchi/wav2vec-heart-sounds}

\newpage
\bibliographystyle{IEEEtranN}
\bibliography{Bibliography.bib}

\end{document}

%% file: styles.tex
\newcommand{\markweighting}[1]
{%
}

\newcommand{\markexplanation}[1]
{%
}

%% file: Figures/ModelProcedure.tex
\begin{tikzpicture}[every node/.style={text width=2.5cm, align=center},auto,node distance=1.5cm]

\usetikzlibrary{positioning, arrows.meta, decorations.markings}

\tikzstyle{vecArrow1} = [thick, decoration={markings,mark=at position
   1 with {\arrow[semithick]{open triangle 60}}},
   double distance=1.4pt, shorten >= 5.5pt,
   preaction = {decorate},
   postaction = {draw,line width=1.4pt, white,shorten >= 4.5pt}]
\tikzstyle{innerWhite} = [semithick, white,line width=1.4pt, shorten >= 4.5pt]
\tikzstyle{vecArrow2} = [thick, decoration={markings,mark=at position 1 with {\arrow[semithick]{triangle 60}}}, double distance=1.4pt, shorten >= 5.5pt, preaction={decorate}, postaction={draw,line width=1.4pt, black,shorten >= 4.5pt}]

\tikzset{%
    block/.style= {rectangle, draw, fill=white!20, minimum height=2em, minimum width=4em},%
    arrow/.style= {thick, ->, >=stealth},%
    line/.style= {thick, -},%
}

\node [block] (sig) {Original Signals};
\node [block, text width=3cm ,below left=0.5cm and 0.5cm of sig.south] (aug) {Generating Augmented Signals};
\node [block, below right=0.5cm and 0.5cm of sig.south] (syn) {Generating  Synthetic Signals};
\node [block, below=2.5cm of sig.south] (pre) {Preprocessing};
\node [block, below of=pre] (train) {Model Training};
\node [block, below of=train] (model) {Trained Model};

\draw [vecArrow1] (sig) -- (pre);
\draw [vecArrow1] (sig) -| (aug);
\draw [vecArrow1] (sig) -| (syn);
\draw [vecArrow1] (aug) |- (pre);
\draw [vecArrow1] (syn) |- (pre);
\draw [vecArrow1] (pre) -- (train);
\draw [vecArrow1] (train) -- (model);

\end{tikzpicture}

%% file: Figures/EPCGAugmentation.tex
\begin{tikzpicture}[every node/.style={text width=3cm, align=center},auto,node distance=1.5cm]

\usetikzlibrary{positioning, arrows.meta, decorations.markings}

\tikzstyle{vecArrow1} = [thick, decoration={markings,mark=at position
   1 with {\arrow[semithick]{open triangle 60}}},
   double distance=1.4pt, shorten >= 5.5pt,
   preaction = {decorate},
   postaction = {draw,line width=1.4pt, white,shorten >= 4.5pt}]
\tikzstyle{innerWhite} = [semithick, white,line width=1.4pt, shorten >= 4.5pt]
\tikzstyle{vecArrow2} = [thick, decoration={markings,mark=at position 1 with {\arrow[semithick]{triangle 60}}}, double distance=1.4pt, shorten >= 5.5pt, preaction={decorate}, postaction={draw,line width=1.4pt, black,shorten >= 4.5pt}]

\tikzset{%
    block/.style= {rectangle, draw, fill=white!20, minimum height=2em, minimum width=4em},%
    arrow/.style= {thick, ->, >=stealth},%
    line/.style= {thick, -},%
}

\node [block] (pcg) {Original PCG Data};
\node [block, right=1cm of pcg.east] (ecg) {Original ECG Data};
\node [block, below of=pcg] (hpss) {HPSS Emphasis Filter};
\node [block, below of=hpss] (noise1p) {White Noise};
\node [block, below of=ecg] (noise1e) {White Noise};
\node [block, below of=noise1e] (wander) {Baseline Wander};
\node [block, below left=0.5cm and 0.5cm of wander.south] (stretch) {Time Stretching};
\node [block, below left=0.5cm and 0.5cm of stretch.south] (am) {Amplitude Modulation};
\node [block, below of=am] (noise2p) {White Noise};
\node [block, right=1cm of noise2p.east] (noise2e) {White Noise};
\node [block, below of=noise2p] (peqp) {Parametric Equalisation (band emphasis)};
\node [block, below of=noise2e] (peqe) {Parametric Equalisation (band emphasis)};
\node [block, below of=peqp] (clinicalp) {Clinical Noise};
\node [block, below of=peqe] (clinicale) {Clinical Noise};
\node [block, below of=clinicalp] (maskp) {Time/Frequency Masking};
\node [block, below of=clinicale] (maske) {Time/Frequency Masking};
\node [block, below of=maskp] (augmentp) {Augmented PCG Data};
\node [block, below of=maske] (augmente) {Augmented ECG Data};

\draw [vecArrow1] (pcg) -- (hpss);
\draw [vecArrow1] (hpss) -- (noise1p);
\draw [vecArrow2] (ecg) -- (noise1e);
\draw [vecArrow1] (noise1p) |- (stretch);
\draw [vecArrow2] (noise1e) -- (wander);
\draw [vecArrow2] (wander) |- (stretch);
\draw [vecArrow1] (stretch) |- (am);
\draw [vecArrow2] (stretch) |- (noise2e);
\draw [vecArrow1] (am) -- (noise2p);
\draw [vecArrow1] (noise2p) -- (peqp);
\draw [vecArrow2] (noise2e) -- (peqe);
\draw [vecArrow1] (peqp) -- (clinicalp);
\draw [vecArrow2] (peqe) -- (clinicale);
\draw [vecArrow1] (clinicalp) -- (maskp);
\draw [vecArrow2] (clinicale) -- (maske);
\draw [vecArrow1] (maskp) -- (augmentp);
\draw [vecArrow2] (maske) -- (augmente);

\end{tikzpicture}

%% file: Figures/mPCGAugmentation.tex
\begin{tikzpicture}[every node/.style={text width=3.2cm, align=center}, node distance=0.6cm]
\usetikzlibrary{positioning, calc, decorations.pathreplacing, arrows.meta, decorations.markings}

\tikzstyle{vecArrow1} = [thick, decoration={markings,mark=at position
   1 with {\arrow[semithick]{open triangle 60}}},
   double distance=1.4pt, shorten >= 5.5pt,
   preaction = {decorate},
   postaction = {draw,line width=1.4pt, white,shorten >= 4.5pt}]
\tikzstyle{innerWhite} = [semithick, white,line width=1.4pt, shorten >= 4.5pt]

\tikzset{
  block/.style = {rectangle, draw, fill=white!20, minimum height=2.1em, minimum width=4.6em},
  faint/.style = {draw=black!40}
}

\node[block] (pcg1)  {Original PCG (Ch 1)};
\node[block, right=5.0cm of pcg1] (pcgN) {Original PCG (Ch n)};
\node (dotsTop) at ($(pcg1)!0.5!(pcgN)$) {$\cdots$}; 

\draw[decorate, decoration={brace, amplitude=6pt}]
  ($(pcg1.north west)+(-0.2,0.2)$) -- ($(pcgN.north east)+(0.2,0.2)$)
  node[midway, yshift=9pt] {$\times~n~\text{channels}$};

\node[block, below=of pcg1] (hpss1) {HPSS Emphasis Filter};
\node[block, below=of hpss1] (wpre1) {White Noise};

\node[block, below=of pcgN] (hpssN) {HPSS Emphasis Filter};
\node[block, below=of hpssN] (wpreN) {White Noise};

\draw[vecArrow1] (pcg1) -- (hpss1);
\draw[vecArrow1] (hpss1) -- (wpre1);

\draw[vecArrow1] (pcgN) -- (hpssN);
\draw[vecArrow1] (hpssN) -- (wpreN);

\coordinate (wpreMid) at ($(wpre1)!0.5!(wpreN)$);

\node[block, below=1.4cm of wpreMid] (stretch) {Time Stretching };

\draw[vecArrow1] (wpre1) -| (stretch);
\draw[vecArrow1] (wpreN) -| (stretch);

\node[block, below=2cm of wpre1.south] (am1) {Amplitude Modulation};
\node[block, below=of am1]  (wpost1) {White Noise};
\node[block, below=of wpost1] (peq1) {Parametric EQ (band emphasis)};
\node[block, below=of peq1] (clin1) {Clinical Noise};
\node[block, below=of clin1] (mask1) {Time/Frequency Masking};
\node[block, below=of mask1] (out1) {Augmented PCG (Ch 1)};

\node[block, below=2cm of wpreN.south] (amN) {Amplitude Modulation};
\node[block, below=of amN]  (wpostN) {White Noise};
\node[block, below=of wpostN] (peqN) {Parametric EQ (band emphasis)};
\node[block, below=of peqN] (clinN) {Clinical Noise};
\node[block, below=of clinN] (maskN) {Time/Frequency Masking};
\node[block, below=of maskN] (outN) {Augmented PCG (Ch n)};

\draw[vecArrow1] (stretch) -| (am1);
\draw[vecArrow1] (stretch) -| (amN);

\foreach \a/\b in {am1/wpost1, wpost1/peq1, peq1/clin1, clin1/mask1, mask1/out1,
                   amN/wpostN, wpostN/peqN, peqN/clinN, clinN/maskN, maskN/outN} {
  \draw[vecArrow1] (\a) -- (\b);
}

\node at ($(out1)!0.5!(outN)$) {$\cdots$};

\end{tikzpicture}

%% file: Figures/SyntheticProcedure.tex
\begin{tikzpicture}[every node/.style={text width=4.5cm, align=center},auto,node distance=1.5cm]

\usetikzlibrary{positioning, arrows.meta, decorations.markings}

\tikzstyle{vecArrow1} = [thick, decoration={markings,mark=at position
   1 with {\arrow[semithick]{open triangle 60}}},
   double distance=1.4pt, shorten >= 5.5pt,
   preaction = {decorate},
   postaction = {draw,line width=1.4pt, white,shorten >= 4.5pt}]
\tikzstyle{innerWhite} = [semithick, white,line width=1.4pt, shorten >= 4.5pt]
\tikzstyle{vecArrow2} = [thick, decoration={markings,mark=at position 1 with {\arrow[semithick]{triangle 60}}}, double distance=1.4pt, shorten >= 5.5pt, preaction={decorate}, postaction={draw,line width=1.4pt, black,shorten >= 4.5pt}]

\tikzset{%
    block/.style= {rectangle, draw, fill=white!20, minimum height=2em, minimum width=4em},%
    arrow/.style= {thick, ->, >=stealth},%
    line/.style= {thick, -},%
}

\node [block, text width=5cm] (ta) {Training Dataset};
\node [block, below of=ta] (preta) {Preprocessing};
\node [block, below of=preta] (card) {Extract Cardiac Cycles};
\node [block, right=1cm of card.east] (icent) {Inference Dataset};
\node [block, below of=icent] (pre2) {Preprocessing};
\node [block, below of=card] (train) {Diffusion Model Training};
\node [block, below right=0.5cm of train.south] (generate) {Generate Synthetic PCG};

\draw [vecArrow1] (ta) -- (preta);
\draw [vecArrow1] (preta) -- (card);
\draw [vecArrow1] (card) -- (train);
\draw [vecArrow1] (train) |- (generate);
\draw [vecArrow1] (icent) -- (pre2);
\draw [vecArrow1] (pre2) |- (generate);

\end{tikzpicture}